\documentclass[a4paper,12pt]{article}
\usepackage{jheppub}
\usepackage[T1]{fontenc}
\usepackage{amsmath}
\usepackage{amssymb}
\usepackage{graphicx}
\usepackage{esint}
\usepackage{epsfig}
\usepackage{amsfonts}
\usepackage{adjustbox,lipsum}

\definecolor{blue}{cmyk}{1,0.9,0,0.3}

\title{\color{blue}Higgs Phenomenology in the Two-Singlet Model}
\author[a]{Amine Ahriche}\author[b,c]{Abdesslam Arhrib}\author[d]{Salah Nasri}
\affiliation[a]{Department of Physics, University of Jijel, PB 98
Ouled Aissa, DZ-18000 Jijel, Algeria.}
\affiliation[b]{D\'{e}partement de Math\'{e}matiques, Facult\'{e}
des Sciences et Techniques, Universit\'{e} Abdelmalek Essaadi, B.
416, Tangier, Morocco.} \affiliation[c]{Institute of Physics,
Academia Sinica, Nankang, Taipei 11529, Taiwan.}
\affiliation[d]{Department of Physics, UAE University, P.O. Box
17551, Al-Ain, United Arab Emirates.} \emailAdd{aahriche@ictp.it,
aarhrib@ictp.it, snasri@uaeu.ae.ac}

\abstract{We study the phenomenology of the Standard Model (SM)
Higgs sector extended by two singlet scalars. The model predicts two
CP-even scalars $h_{1,2}$ which are a mixture of doublet and singlet
components as well as a pure singlet scalar $S_{0}$ which is a dark
matter candidate. We show that the model can satisfy the relic
density and direct detection constraints as well as all the recent
ATLAS and CMS measurements. We also discuss the effect of the extra
Higgs bosons on the different Higgs triple couplings
$h_{i}h_{j}h_{k}$, $i,j,k=1,2$. A particular attention is given to
the triple self-coupling of the SM-like Higgs where we found that
the one loop corrections can reach 150\% is some cases. We also
discuss some production mechanisms for $h_{1}$ and $h_{2}$ at the
LHC as well as at the future International Linear Collider. It is
found that the production cross section of a pair of SM-like Higgs
bosons could be much larger than the corresponding one in the SM and
would reveal physics beyond the SM if observable. We also show that
in this model the branching ratio of the SM-like Higgs decaying to
two singlet scalars could be of the order of 20\%, therefore the
production of the SM Higgs followed by its decay to a pair of
singlets would be an important source of production of singlet
scalars.} \keywords{Higgs decay, dark matter, singlets.}

\begin{document}

\maketitle

\flushbottom

\section{Introduction}

The Large Hadron Collider (LHC) at CERN has just successfully finished its
first phase of operation with a 7 and 8 TeV run. Both experiments ATLAS and
CMS at the LHC announced last July the discovery of a Higgs-like particle with
a mass in the range 125-126 \textrm{GeV} \cite{ATLAS,CMS}. Both
collaborations, ATLAS and CMS reported a clear excess in the two photon
channel and in the $ZZ^{\ast}$ channel \cite{ATLAS,CMS}. The discovery is also
confirmed with less significance in other channels
\cite{atlasmoriond,cmsmoriond}, like $WW^{\ast}$ which has a lower mass
resolution, and also by the final Tevatron results reported by CDF and D0
experiments \cite{tevatron}.

The extraction of the couplings of the Higgs-like particle to gauge bosons and
fermions achieved up to now from the $7 \oplus8$ TeV data shows that this
particle looks more and more like the SM Higgs boson
\cite{atlasmoriond,cmsmoriond}, while more data is needed in order to fully
pin down the exact nature of the newly discovered particle.

Although, ATLAS and CMS data show no significant deviation of the signal from
the SM predictions. At ATLAS, the diphoton channel shows some small
enhancement. The overall signal strength for diphoton is about $1.55_{-0.28}%
^{+0.33}$, which corresponds to about 2 $\sigma$ deviation from the SM
prediction \cite{ATLAS2G}; while the other channels are consistent with the
SM. However, at CMS, the new analysis for diphoton mode based on multivariate
analysis \cite{CMS2G} gives $0.77\pm0.27$, which is compatible with the SM.
Many models beyond the SM have been proposed to explain the diphoton excess,
but the actual disagreement between ATLAS and CMS does not allow to extract
significant conclusions.

Since the Higgs-like particle decays to two photons, it can not be spin one
particle because of the Young Landau theorem, it is either spin-0 or spin-2.
Recently, spin and parity of the Higgs-like particle were studied from the
angular distribution of the diphoton, $ZZ^{\ast}$ and $WW^{\ast}$ decay
channels \cite{atlaspin,cmspin} at ATLAS and CMS. Both collaborations disfavor
the pure pseudoscalar hypothesis $J^{P}=0^{-}$; and also a pure spin-2
hypothesis. In addition, the spin one hypotheses is also disfavored with an
even higher confidence.

Therefore, the first phase of the LHC run is just the beginning of a precise
measurement program that starts with $7\oplus8$ TeV data and will be completed
with the second run of the LHC at 13-14 TeV as well as by the International
Linear Collider (ILC). It is well known that the precise measurement programs
at the ILC and the LHC are complementary \cite{Weiglein:2004hn,Peskin:2012we}.
Such measurements, if accurate enough, can be also helpful in discriminating
between models through their sensitivity to radiative correction effects, in
particular in specific cases like the decoupling limit. It is well known that
many SM extensions such as SUSY models or extended Higgs sector models possess
such decoupling limit where the light Higgs boson completely mimics the SM Higgs.

ATLAS and CMS discovery, has lead to several phenomenological constraints on
the scalar sector in such extensions of SM Higgs sector with extra doublets,
Higgs sector with doublet and singlets, or Higgs sector with doublet and
triplets etc... The fact that the Higgs-like particle couplings to gauge
bosons and fermions are consistent with the SM predictions; can put severe
constraints on all beyond SM extensions that try to accommodate such
Higgs-like particle.

The aim of this paper is to study the phenomenology of the SM Higgs
sector extended by two real, spinless and $\mathbb{Z}_{2}$ symmetric
fields which can explain the Dark Matter (DM) \cite{TSM,last}. The
model has three CP-even scalars, two of which, $h_{1,2}$, are mixing
of a $SU(2)_{L}$ doublet and a singlet, whereas a
$\mathbb{Z}_{2}$-odd singlet $S_{0}$ remains unmixed, which can play
the role of DM candidate. However, both $h_{1}$ and $h_{2}$ can
decay to a pair of $S_{0}$, if kinematically allowed, it will
contribute to the invisible decay of $h_{1}$ or $h_{2}$; and will
potentially modify the properties of the Higgs-like particle $h_{1}$
or $h_{2}$. In addition, the annihilation of $S_{0}$ into SM
particles will provide thermal relic density and the scattering of
$S_{0}$ on nucleons will lead to direct detection signatures.

In the light of the recent discovery of a 125 \textrm{GeV}
Higgs-like particle \cite{ATLAS,CMS}, we investigate, in the
framework of the two singlets model, the possibility that one of the
scalars $h_{1}$ or $h_{2}$ is the particle observed by ATLAS and
CMS. Therefore, we consider the two cases where one of the scalar
eigenmasses $m_{1}$ or $m_{2}$ lies in the range 123.5-127.5
\textrm{GeV} tolerated by ATLAS and CMS experimental results, with
their couplings to the SM fermions and gauge bosons close to the SM
case, i.e.,
$g_{h_{i}f\bar{f}}^{2}/g_{hf\bar{f}}^{2(SM)}=g_{h_{i}VV}^{2}/g_{hVV}^{2(SM)}\geq0.9$.
Then, we will investigate the phenomenology of the non SM-like Higgs
in both cases.

This paper is organized as follows. We first introduce the two singlet model
and its theoretical constraints in the second section. We investigate the DM
and its direct detection constraints on the two singlet model in the third
section. Section IV is devoted to various Higgs triple self-couplings that
exist in this model with particular attention given to the triple
self-coupling of the SM-like Higgs scalar. We discuss some phenomenological
aspects of the model such as the Higgs decays and double Higgs production in
section V and present our conclusion in section VI. In the appendices, we give
the tree-level cubic and quartic scalar couplings and we provide the details
of the calculation of the effective Higgs triple couplings from the effective potential.

\section{The Two-Singlet Model}

In this model, we extend the Standard Model with two real scalar fields
$S_{0}$ and $\chi_{1}$; which transform under the discrete symmetry
$\mathbb{Z}_{2}^{(0)}\otimes\mathbb{Z}_{2}^{(1)}$ as%
\begin{equation}%
\begin{array}
[c]{ccc}%
\mathbb{Z}_{2}^{(0)}: &  & (S_{0},\chi_{1})\rightarrow(-S_{0},\chi_{1})\\
\mathbb{Z}_{2}^{(1)}: &  & (S_{0},\chi_{1})\rightarrow(S_{0},-\chi_{1}).
\end{array}
\end{equation}
The field $\chi_{1}$ has a non vanishing vacuum expectation value,
which breaks $\mathbb{Z}_{2}^{(1)}$ spontaneously, whereas,
$\left\langle S_{0}\right\rangle =0$; and hence, $S_{0}$ is a dark
matter candidate. Both fields are standard model gauge singlets and
hence can interact with 'visible'\ particles only via the Higgs
doublet $H$. The part of the Lagrangian that includes the fields
$S_{0}$, $H$, and $\chi_{1}$ is written as follows:
\begin{equation}
\mathcal{L}=\left(  D_{\mu}H\right)  ^{\dag}D_{\mu}H+\frac{1}{2}\left(
\partial_{\mu}S_{0}\right)  +\frac{1}{2}\left(  \partial_{\mu}\chi_{1}\right)
-V(H,\chi_{1},S_{0}),
\end{equation}
with
\begin{align}
H^{T} &  =\left(  h^{+},~~(\upsilon+\widetilde{h}+i\chi_{0})/\sqrt{2}\right)
,~D_{\mu}H=\left(  \partial_{\mu}-ig_{2}/2\sigma^{a}W_{\mu}^{a}-ig_{1}%
/2B_{\mu}\right)  H,\nonumber\\
\chi_{1} &  =\upsilon_{1}+\widetilde{\chi}_{1},
\end{align}
where $\sigma^{a}$ are the Pauli matrices, $W_{\mu}^{a}$ ($B_{\mu}$) and
$g_{2}$ ($g_{1}$) are the $SU(2)_{L}$ ($U(1)_{Y}$) gauge field and coupling,
respectively. The tree-level scalar potential that respects the $\mathbb{Z}%
_{2}$ symmetries is given by \cite{TSM}
\begin{align}
V(H,\chi_{1},S_{0}) &  =-\mu^{2}H^{\dag}H+\frac{\lambda}{6}\left(  H^{\dag
}H\right)  ^{2}+\frac{\widetilde{m}_{0}^{2}}{2}S_{0}^{2}-\frac{\mu_{1}^{2}}%
{2}\chi_{1}^{2}+\frac{\eta_{0}}{24}S_{0}^{4}+\frac{\eta_{1}}{24}\chi_{1}%
^{4}\nonumber\\
&  +\frac{\lambda_{0}}{2}S_{0}^{2}H^{\dag}H+\frac{\lambda_{1}}{2}\chi_{1}%
^{2}H^{\dag}H+\frac{\eta_{01}}{4}S_{0}^{2}\chi_{1}^{2}.\label{TLP}%
\end{align}
The parameters $\mu^{2}$ and $\mu_{1}^{2}$\ could be eliminated from the
potential by imposing ($\upsilon,\upsilon_{1}$) to be the absolute minimum as
\begin{align}
\mu^{2} &  =\lambda\upsilon^{2}/6+\lambda_{1}\upsilon_{1}^{2}/2+\tfrac
{1}{\upsilon}\left.  \tfrac{\partial}{\partial\tilde{h}}V^{1-l}\right\vert
_{\tilde{h}=\upsilon,\chi_{1}=\upsilon_{1},S_{0}=0.},\nonumber\\
\mu_{1}^{2} &  =\eta_{1}\upsilon_{1}^{2}/6+\lambda_{1}\upsilon^{2}/2+\tfrac
{1}{\upsilon_{1}}\left.  \tfrac{\partial}{\partial\chi_{1}}V^{1-l}\right\vert
_{\tilde{h}=\upsilon,\chi_{1}=\upsilon_{1},S_{0}=0.},\label{mss}%
\end{align}
where $V^{1-l}$ is the one-loop corrections to the scalar potential. While the
condition%
\begin{equation}
\widetilde{m}_{0}^{2}+\lambda_{0}\upsilon^{2}/2+\eta_{01}\upsilon_{1}%
^{2}/2+\left.  \tfrac{1}{S_{0}}\tfrac{\partial}{\partial S_{0}}V^{1-l}%
\right\vert _{\widetilde{h}=\upsilon,\chi_{1}=\upsilon_{1},S_{0}%
=0.}>0,\label{s0d}%
\end{equation}
should be fulfilled in order that the potential does not develop a vev in the
direction of $S_{0}$. In fact, the conditions (\ref{s0d}) are not enough to
guaranty the vacuum being ($\upsilon,\upsilon_{1}$); one must require that the
Jacobian must be positive, which is equivalent to the fact that the two
mass-squared eigenvalues are positive. In addition, we impose the vacuum
stability condition
\begin{equation}
\allowbreak\lambda\eta_{0}\eta_{1}-9\eta_{0}\lambda_{1}^{2}-9\lambda\eta
_{01}^{2}-9\eta_{1}\lambda_{0}^{2}+54\lambda_{0}\lambda_{1}\eta_{01}%
>0,\label{sV}%
\end{equation}
where $\lambda$, $\eta_{1}$ and $\eta_{0}$ must be strictly positive, while
$\lambda_{0}$, $\lambda_{1}$ and $\eta_{01}$\ could have negative values
within the condition (\ref{sV}). Moreover, $\lambda$, $\eta_{1}$ $\eta_{0}$,
$\lambda_{0}$, $\lambda_{1}$\ and $\eta_{01}$ must remain perturbative.

The spontaneous breaking of the electroweak and the $\mathbb{Z}_{2}$
symmetries introduces the two vacuum expectation values $\upsilon$ and
$\upsilon_{1}$ respectively. With the value of $\upsilon$ being fixed
experimentally to 246 \textrm{GeV} from W gauge boson mass, the model has ten
parameters. The minimization conditions of the effective potential allows one
to eliminate $\mu^{2}$ and $\mu_{1}^{2}$\ in favor of $(\upsilon,\upsilon
_{1})$. Then, we are left with eight parameters: $\lambda$, $\lambda_{0}$,
$\lambda_{1}$, $\eta_{0}$, $\eta_{1}$, $\eta_{01}$, $\upsilon_{1}$ and $m_{0}%
$. However, the DM self-coupling constant $\eta_{0}$ does not enter the
calculations of the lowest-order processes of this work, so effectively, we
are left with seven input parameters.

The physical Higgs scalars $h_{1}$ and\ $h_{2}$, with masses $m_{1}$ and
$m_{2}$ (with $m_{1}<m_{2}$), are related to the excitations of the neutral
component of the SM Higgs doublet field, Re$(H^{(0)})=(\upsilon+{\widetilde
{h}})\sqrt{2}$, and the field $\chi_{1}=\widetilde{\chi}_{1}+\upsilon_{1}$
through a mixing angle $\theta$. The scalars $\widetilde{h}$ and
$\widetilde{\chi}_{1}$\ are not the interacting fields but components of the
eigenstates $h_{1}$ and $h_{2}$ which are obtained after the electroweak and
the $\mathbb{Z}_{2}$ symmetries are spontaneously broken. Then the
interactions of the DM candidate with the scalar sector that is relevant to
the relic density, are not these in (\ref{TLP}), but instead, their
modification
\begin{equation}
\left(
\begin{array}
[c]{c}%
h_{1}\\
h_{2}%
\end{array}
\right)  =\left(
\begin{array}
[c]{cc}%
\cos\theta & \sin\theta\\
-\sin\theta & \cos\theta
\end{array}
\right)  \left(
\begin{array}
[c]{c}%
\widetilde{h}\\
\widetilde{\chi}_{1}%
\end{array}
\right)  , \label{mix}%
\end{equation}
as shown in (\ref{Vssb}). In our work, the CP-even scalar masses and the
mixing angle are estimated at one-loop. Here the quartic interactions get
modified and new cubic interactions emerge \cite{TSM}. The couplings of the
$h_{1}$ and $h_{2}$ with fermions and gauge fields are just the projections of
the doublets couplings using (\ref{mix}). The scalar potential that emerges
after the electroweak symmetry breaking is given as a function of scalar
eigenstates by%
\begin{align}
V(h_{1},h_{2},S_{0})  &  =\frac{m_{0}^{2}}{2}S_{0}^{2}+\frac{m_{1}^{2}}%
{2}h_{1}^{2}+\frac{m_{2}^{2}}{2}h_{2}^{2}\nonumber\\
&  +\frac{\lambda_{001}^{(3)}}{2}S_{0}^{2}h_{1}+\frac{\lambda_{002}^{(3)}}%
{2}S_{0}^{2}h_{2}+\frac{\lambda_{111}^{(3)}}{6}h_{1}^{3}+\frac{\lambda
_{222}^{(3)}}{6}h_{2}^{3}+\frac{\lambda_{112}^{(3)}}{2}h_{1}^{2}h_{2}%
+\frac{\lambda_{122}^{(3)}}{2}h_{1}h_{2}^{2}\nonumber\\
&  +\frac{\eta_{0}}{24}S_{0}^{4}+\frac{\lambda_{1111}^{(4)}}{24}h_{1}%
^{4}+\frac{\lambda_{2222}^{(4)}}{24}h_{2}^{4}+\frac{\lambda_{0011}^{(4)}}%
{4}S_{0}^{2}h_{1}^{2}+\frac{\lambda_{0022}^{(4)}}{4}S_{0}^{2}h_{2}^{2}%
+\frac{\lambda_{0012}^{(4)}}{2}S_{0}^{2}h_{1}h_{2}\nonumber\\
&  +\frac{\lambda_{1112}^{(4)}}{6}h_{1}^{3}h_{2}+\frac{\lambda_{1122}^{(4)}%
}{4}h_{1}^{2}h_{2}^{2}+\frac{\lambda_{1222}^{(4)}}{6}h_{1}h_{2}^{3},
\label{Vssb}%
\end{align}
where the triple and quartic coupling are given in appendix A. In our analysis
we require that:

\begin{itemize}
\item[(i)] all the dimensionless quartic couplings to be $\ll4\pi$ for the
theory to remain perturbative,

\item[(ii)] they have to be chosen in such a way that the ground state
stability is insured;

\item[(iii)] and we assume that the DM mass lies up to 1 TeV.
\end{itemize}

In our work, we consider the following values for the free parameters;
\begin{align}
&  \lambda,\eta_{0},\eta_{1},\left\vert \lambda_{0}\right\vert ,\left\vert
\lambda_{1}\right\vert ,\left\vert \eta_{01}\right\vert <3\nonumber\\
&  20<\frac{\upsilon_{1}}{\mathrm{GeV}}<2000,~1<\frac{m_{0}}{\mathrm{GeV}%
}<1000, \label{Reg}%
\end{align}
and we make random choices taking into account the value of the relic density
lying in the physical interval (\ref{Omega}) and being not in conflict with
direct detection DM experiments. Also, one of the CP even scalars mass lies
around 123.5-127.5 $\mathrm{GeV}$, with couplings to SM fermions and gauge
bosons that are similar to the SM by more than $\epsilon\gtrsim$ 90\%, where
$\epsilon$ is $\cos^{2}\theta$ or $\sin^{2}\theta$ depending if $h_{1}$ or
$h_{2}$ is the SM-like Higgs, respectively.

For our numerical illustration, we define the following two
scenarios: A and B where the SM-like Higgs is $h_{1}$ and $h_{2}$
respectively. In addition, the invisible decay channel in case A
$h_{1}\rightarrow2DM$ could be open up to 20\%, while both
$h_{2}\rightarrow2DM$ and $h_{2}\rightarrow h_{1}h_{1}$ should not
exceed together 20\% in case B. In fact, the former constraint on
the invisible decay originates from global fit analysis to ATLAS and
CMS data \cite{global,global0,strumia}. When deriving this limit in
a global analysis, it is assumed that the Higgs boson has similar
couplings to fermions and gauge bosons as in the SM and additional
invisible decay modes. For instance, if the effective
gluon-gluon-Higgs, $\gamma$-$\gamma$-Higgs or Higgs couplings to
fermions are considered, the Higgs couplings to gauge bosons are
modified, and therefore the above limit could be exceeded
\cite{global0,strumia}. Therefore, in our work, we consider the
conservative choice $B(h\rightarrow invisible)\leq20\%$. Recently,
both ATLAS and CMS have searched for invisible decay of the Higgs.
Assuming the Higgs-strahlung SM cross section for $pp\rightarrow ZH$
with 125 \textrm{GeV} SM Higgs boson, ATLAS exclude with 95\%
confidence level an invisible branching fraction of the Higgs larger
than 65\% and CMS obtain similar result \cite{Higgs-invisible}. CMS
also looks for invisible decay of the Higgs through vector boson
fusion process and exclude an invisible branching fraction of the
Higgs larger than 69\% \cite{CMS-VBF}. When data from $pp\rightarrow
ZH$ and VBF are combined the limit becomes 54\% \cite{CMS-VBF}.

In our numerical scans, we will consider the parameter values that:

\begin{itemize}
\item ensure that one CP-even scalar is the SM-like by more than 90\%,

\item give the right amount of the DM relic density,

\item do not conflict the direct detection DM experiments such as CDMSII
\cite{CDMSII} and Xe100 \cite{Xe},

\item in case A, the heavy scalar $h_{2}$ escapes the ATLAS \cite{ATLAS} and
CMS \cite{CMS} bounds; and in case B, the light Higgs escapes the LEP
constraints \cite{OPAL};

\item and the invisible SM-like Higgs decay channel should not exceed 20\%.
\end{itemize}

\section{Dark Matter \& Detection}

In the framework of the thermal dynamics of the Universe within the standard
cosmological model \cite{kolb}, the WIMP relic density is related to its
annihilation rate by the familiar relation:
\begin{equation}
\Omega_{D}\bar{h}^{2}=\frac{1.07\times10^{9}x_{f}}{\sqrt{g_{\ast}}%
m_{Pl}\left\langle \upsilon_{12}\sigma_{ann}\right\rangle \mathrm{GeV}},
\label{Om}%
\end{equation}
with
\begin{equation}
x_{f}=\ln\frac{0.038~m_{Pl}m_{0}\left\langle \upsilon_{12}\sigma
_{ann}\right\rangle }{\sqrt{g_{\ast}x_{f}}}.
\end{equation}
The notations are as follows: the quantity $\bar{h}$ is the Hubble constant in
units of 100 $km\times s^{-1}\times Mpc^{-1}$, the quantity $m_{Pl}%
=1.22\times10^{19}$ \textrm{GeV} the Planck mass, $m_{0}$ the DM mass,
$x_{f}=m_{0}/T_{f}$ the ratio of the DM mass to the freeze-out temperature
$T_{f}$ and $g_{\ast}$ the number of relativistic degrees of freedom with mass
less than $T_{f}$. The quantity $\left\langle \upsilon_{12}\sigma
_{ann}\right\rangle $ is the thermally averaged annihilation cross section of
a pair of two DM particles multiplied by their relative velocity in the
center-of-mass reference frame \cite{TSM}. When considering the current value
for the DM relic density \cite{DM}
\begin{equation}
\Omega_{D}\bar{h}^{2}=0.1187\pm0.0017; \label{Omega}%
\end{equation}
and taking the approximate values of $x_{f}\approx19.2\sim21.6$ and
$m_{0}\approx10\sim100$ $\mathrm{GeV}$, we get%
\begin{equation}
\left\langle \upsilon_{12}\sigma_{ann}\right\rangle =(1.9\pm0.2)\times
10^{-9}\text{ }\mathrm{GeV}^{-2}. \label{sig}%
\end{equation}
The value in (\ref{sig}) for the DM annihilation cross section translates into
a relation between the parameters of a given theory entering the calculated
expression of $\left\langle \upsilon_{12}\sigma_{ann}\right\rangle $, hence
imposing a constraint on these parameters will limit some of the possible
range of DM masses. These constraints can be exploited to examine aspects of
the theory like perturbativity, while at the same time reducing the number of
parameters by one. However, since we will consider a wide range for the DM
mass, $1\sim1000 $ $\mathrm{GeV}$, the ratio $x_{f}$ will be estimated
numerically using (\ref{Om}), especially for small mass values. Depending on
how heavy/light is the DM candidate, its main annihilation channel will be to
fermion pairs $f\bar{f}$ ($b\bar{b}$, $c\bar{c}$, $\tau\bar{\tau}$, or
$\mu\bar{\mu}$), but for very large mass values, the channels $h_{1}h_{1}$,
$h_{1}h_{2}$, $h_{2}h_{2}$, $WW$, $WW^{\ast}$, $ZZ$, $ZZ^{\ast}$ and $t\bar
{t}$ could be also important. All the explicit formula of the annihilation
cross section are given in \cite{TSM}.

During previous years, experiments such as CDMS II \cite{CDMSII}, XENON 10/100
\cite{Xe} and CoGeNT \cite{CoGe} have been searching for signal of elastic
scattering of a DM WIMP off nucleon targets in deep underground. Although, no
unambiguous signal has been seen yet, they yielded increasingly stringent
exclusion bounds on the DM-nucleon elastic scattering total cross section
$\sigma_{det}$ in terms of the DM mass $m_{0}$. The direct detection cross
section for the scattering of $S_{0}$ (the DM candidate in this model) off
nucleon, $\sigma_{det}$, is given by \cite{TSM, last}%
\begin{equation}
\sigma_{det}=\frac{g_{HNN}^{2}m_{N}^{2}}{4\pi(m_{N}+m_{0})^{2}}\left[
\frac{\lambda_{001}^{(3)}\cos\theta}{m_{1}^{2}}-\frac{\lambda_{002}^{(3)}%
\sin\theta}{m_{2}^{2}}\right]  ^{2}, \label{Sigdet}%
\end{equation}
where $m_{N}$ is the nucleon mass, $\lambda_{00i}^{(3)}$ is the coupling
constants of $h_{i}S_{0}^{2}$ given in appendix A, and $g_{HNN}$ is the
effective Higgs-nucleon coupling\textbf{, }which is estimated based on heavy
baryon chiral perturbation theory to be $g_{HNN}\simeq1.5\times10^{-3}%
~$\cite{SVZ, Ch, GLS}, whereas lattice calculations give somehow smaller
values \cite{Lat1, Lat2}.

\begin{figure}[h]
\begin{center}
\includegraphics[width=8cm,height=6cm]{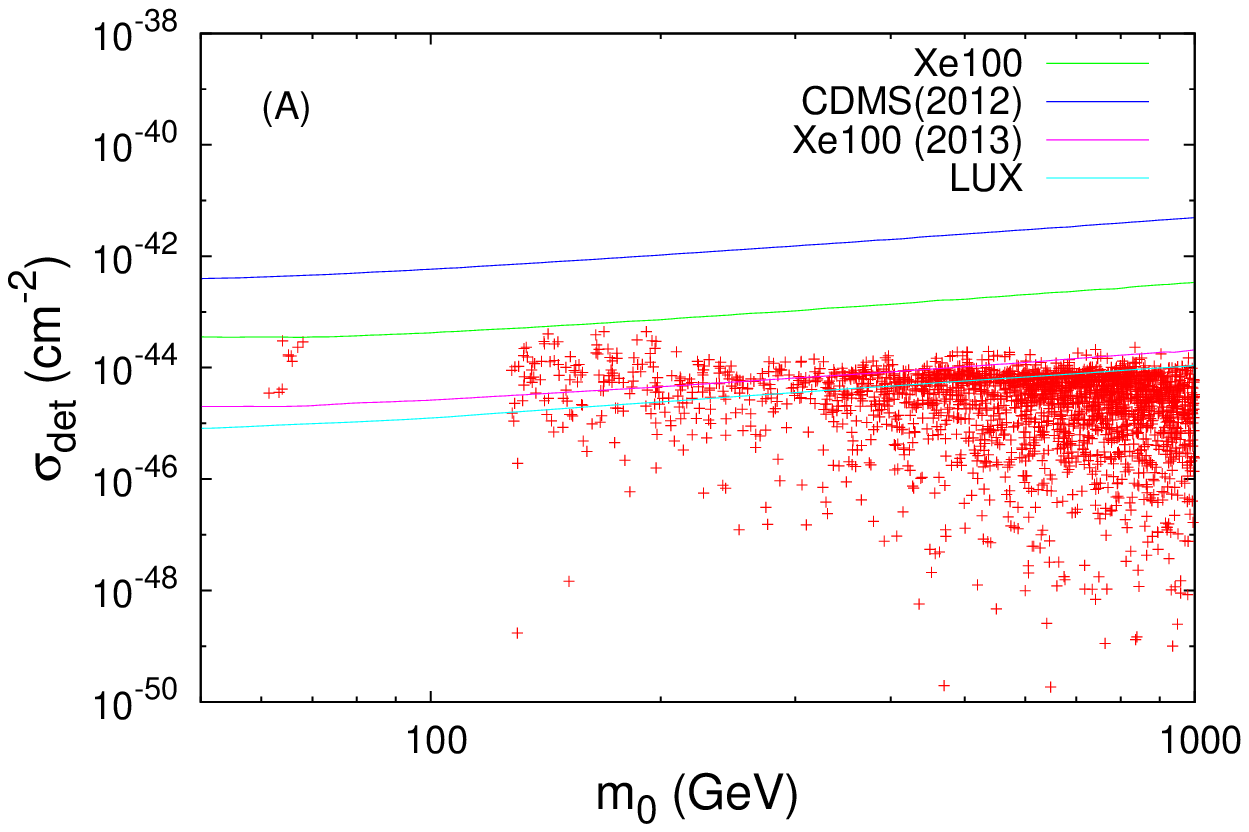}~\includegraphics[width=7.6cm,height=5.8cm]{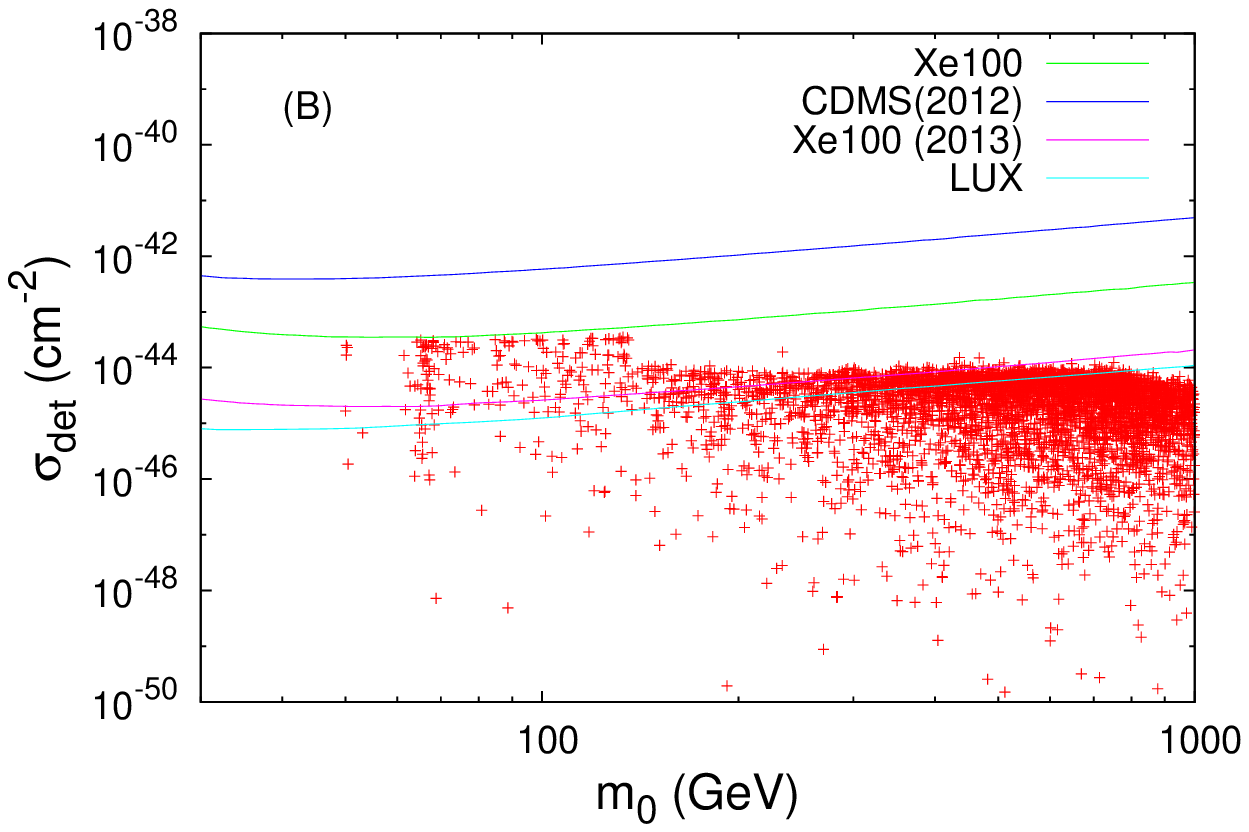}
\end{center}
\caption{\textit{The direct detection cross section versus the DM
mass compared to the recent Xe100 and LUX results, where the left
and right panels correspond to the cases A and B respectively. It is
clear that all the considered benchmarks are not in conflict with
previous experimental
bounds such as Xe100 (2012) and CDMSII (2012).}}%
\label{Sig}%
\end{figure}

In our work, the free parameters are chosen in such a way that the
spectrum of the scalar sector has a SM Higgs like particle of 125
\textrm{GeV}, and the relic density of $S_{0}$ is consistent with
the Planck data \cite{DM}. As it is shown in Fig. \ref{Sig}, we find
that for most of the benchmarks, the elastic scattering cross
section $\sigma_{det}$ is below $10^{-45}$ $cm^{2}$, i.e., below all
the experimental bounds including the new one from Xe100 as well the
latest LUX results \cite{LUX}, especially for DM masses larger than
125 \textrm{GeV} for case A; and 50 \textrm{GeV} for case B.

This behavior could be due to the cancelation between the two terms inside the
bracket in Eq. (\ref{Sigdet}) or/and to the scaling of $\sigma_{det}$ as the
inverse square of $m_{0}$ which results in the suppression of the heavy DM
event rate. However, for DM lighter than $30$~\textrm{GeV}, the invisible
Higgs decay fraction exceeds $20\%$, and so it is in conflict with ATLAS and
CMS data.

\section{The Triple Higgs Coupling}

With the discovery of the Higgs-like particle at ATLAS and CMS with a mass in
the range 125-126 \textrm{GeV}, and in order to establish the Higgs mechanism
for the electroweak symmetry breaking we need to measure not only Higgs
couplings to fermions and gauge bosons but also the triple and quartic
self-coupling of the Higgs boson which are necessary for Higgs potential
reconstruction. The measurement of the triple and quartic couplings, if
precise enough, can help distinguishing between various SM extensions. The
Triple Higgs self-coupling can be, in principle, measured directly in
pair-production of Higgs boson at the LHC with high luminosity option
\cite{double1} and/or at $e^{+}e^{-}$ International Linear Collider
\cite{Weiglein:2004hn}. \newline At the LHC, it is rather difficult to
reconstruct the triple coupling of the Higgs because of the smallness of the
cross section $gg\rightarrow hh$ as well as the large associated QCD
background. Several parton level analysis have been devoted to this process
with the following final states: $hh\rightarrow W^{+}W^{-}W^{+}W^{-}$ (which
would lead to same sign leptons) \cite{double2}, $hh\rightarrow b\bar{b}%
W^{+}W^{-}$ \cite{double0}, $hh\rightarrow b\bar{b}\gamma\gamma$
\cite{double3} and $hh\rightarrow b\bar{b}\tau^{+}\tau^{-}$ \cite{double0}.
The last two processes seem to be very promising for High luminosity at the
LHC. The authors in ref.~\cite{Dolan:2012rv} used the recent jet substructure
techniques to study the Higgs pair production and the Higgs pair production in
association with hard jet, where it is found that $b\bar{b}\tau^{+}\tau^{-}$
and $b\bar{b}\tau^{+}\tau^{-}+jet$ channels can be used to constrain the
Triple Higgs self-coupling in the SM.

On the other hand, at the ILC, the process $e^{+}e^{-}\rightarrow
Zhh\rightarrow l^{+}l^{-}b\bar{b}b\bar{b}$ has been investigated with 500
\textrm{GeV} center of mass energy with 1 ab$^{-1}$ luminosity and it turns
out that this process can be useful for measuring the Higgs self-coupling at
the ILC \cite{Weiglein:2004hn}.

In our study, the Triple Higgs self-couplings are estimated by taking the
third derivatives of the effective potential at one-loop using the exact
formulae given in appendix B, where we show how the renormalization scale
disappears in favor of measured quantities. In our model, the deviation of the
Triple Higgs self-coupling from the SM value can not come only from the
modification of Higgs couplings to top quarks through the reduction factor
$\epsilon$ (see Eq. (\ref{EP})), but also comes from new contributions of the
other Higgs scalar and the DM candidate. In the rest of this section, for both
cases A and B, we estimate the magnitude of different scalar triple couplings
at one-loop and their deviation from the SM value. In what follows, the
renormalization scale is taken to be the Higgs mass 125 \textrm{GeV}.

\subsection*{Case A: $h_{1}$ SM-like}

In this case, $h_{1}$ is the SM-like while $h_{2}$ is dominated by singlet
component. The relevant Triple Higgs self-couplings are $\lambda_{h_{1}%
h_{1}h_{1}}$, $\lambda_{h_{1}h_{1}h_{2}}$ and $\lambda_{h_{1}h_{2}h_{2}}$,
where the first one corresponds to $\lambda_{hhh} $ {\textbf{i}n the SM case}.
The other two couplings $\lambda_{h_{1}h_{1}h_{2}}$ and $\lambda_{h_{1}%
h_{2}h_{2}}$ have at least one $h_{1}$ leg which could give access to an
associate production $h_{2}h_{1}$ or double production $h_{2}h_{2}$ through:
$pp\rightarrow h_{1}^{\ast}\rightarrow h_{1}h_{2}$, $pp\rightarrow h_{1}%
^{\ast}\rightarrow h_{2}h_{2}$ at the LHC or $e^{+}e^{-}\rightarrow
Zh_{1}^{\ast}\rightarrow Zh_{1}h_{2}$, $e^{+}e^{-}\rightarrow Zh_{1}^{\ast
}\rightarrow Zh_{2}h_{2}$ at the ILC.

\begin{figure}[h]
\begin{center}
\includegraphics[width=8cm,height=6cm]{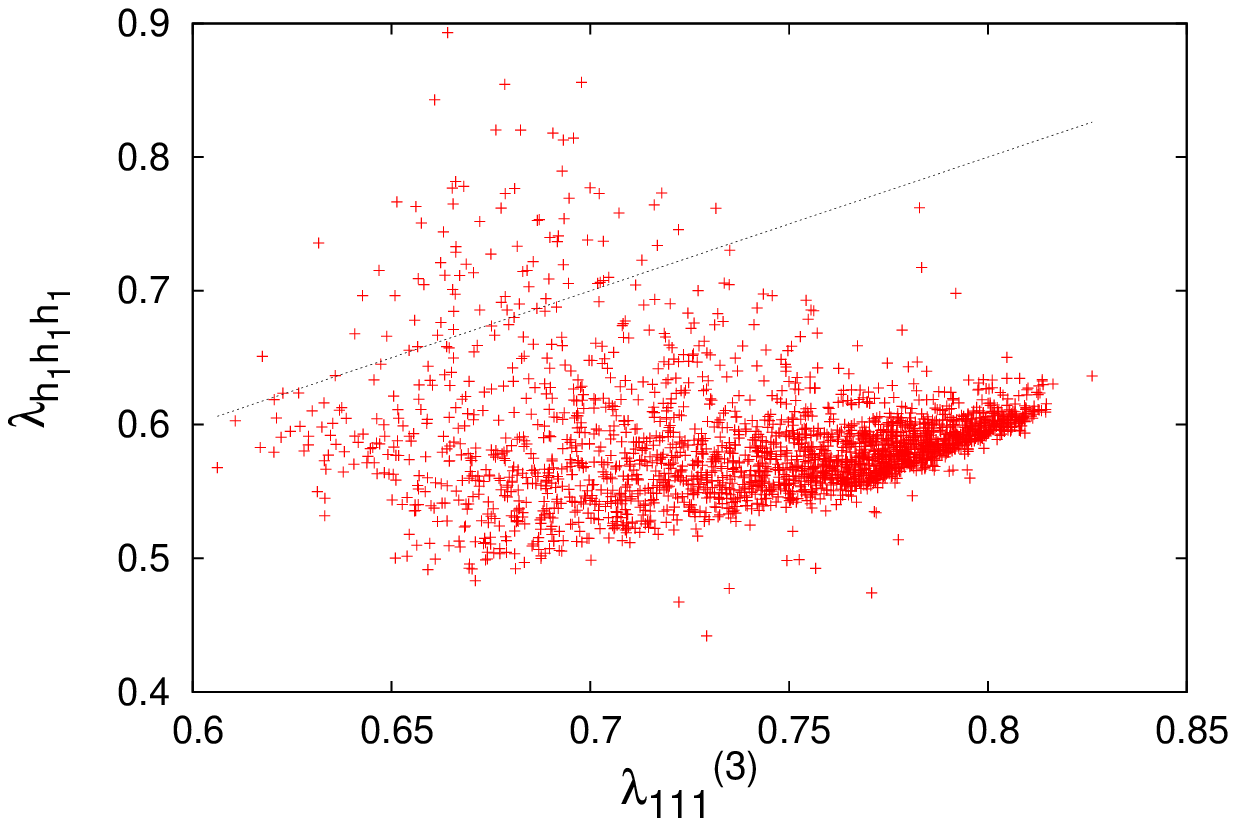}\includegraphics[width=8cm,height=6cm]{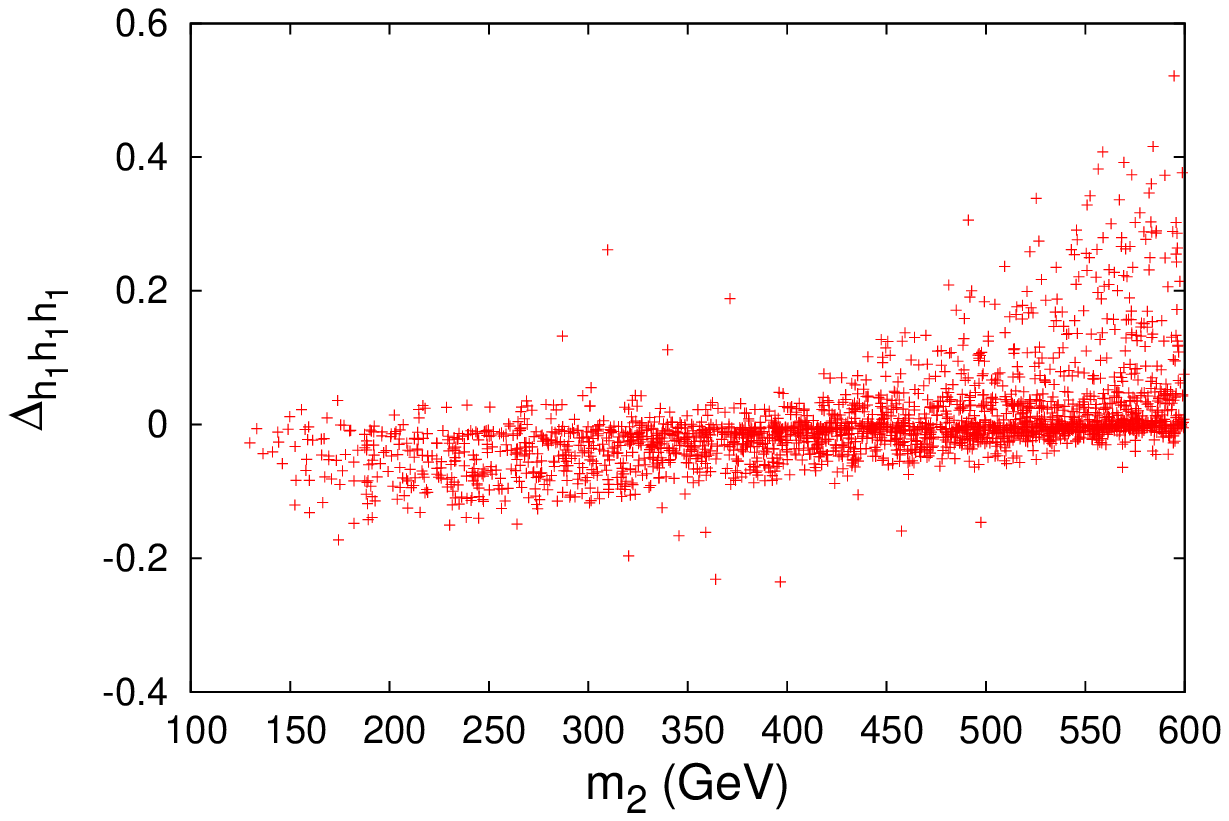}
\end{center}
\caption{\textit{Left panel: the SM-like triple Higgs coupling
versus its tree-level value in units of the EW vev value for
randomly chosen sets of parameters, where the SM-like Higgs is
defined according to case A. Right panel: the relative enhancement
on the SM-like triple Higgs coupling of with respect to the SM value
versus the mass of the heavy scalar $m_{2}$ for
the same sets of parameters.}}%
\label{D1}%
\end{figure}

In order to illustrate the magnitude of the one-loop corrected triple Higgs
couplings, we show in Fig. \ref{D1}-left the triple SM-like Higgs coupling
versus its tree-level value. It is clear that only the coupling $\lambda
_{h_{1}h_{1}h_{1}}$ which receives significant corrections at the one-loop
level and make it larger than its corresponding tree level value. Also one has
to mention that its value is the smallest one with respect to the other ones:
$\lambda_{h_{1}h_{2}h_{2}}$ and $\lambda_{h_{1}h_{1}h_{2}}$. Note that the
value of $\lambda_{h_{2}h_{2}h_{2}}$ (which is not shown here) could be much
larger than the others, i.e. up to $\lambda_{h_{2}h_{2}h_{2}}\sim
8\times\upsilon$.

In order to show the effect of these new contributions on this triple coupling
$\lambda_{h_{1}h_{1}h_{1}}$, we define the following quantity $\Delta
_{h_{1}h_{1}h_{1}}=(\lambda_{h_{1}h_{1}h_{1}}-\lambda_{hhh}^{SM}%
)/\lambda_{hhh}^{SM}$, which represents the relative enhancement on the triple
Higgs coupling at one-loop with respect to the same quantity estimated at
one-loop in the SM for the recently measured Higgs mass. 

In Fig.~\ref{D1}-right, we show $\Delta_{h_{1}h_{1}h_{1}}$ as a function of
the heavy scalar mass $m_{2}$. It is clear that in this case, the one-loop
corrections to the SM-like Higgs $h_{1}$ could have an enhancement greater
than\ 40\%. Since we have subtracted the SM contribution at one-loop, this
enhancement is then attributed to the new contributions of $h_{2}$ and $S_{0}
$.

\subsection*{Case B: $h_{2}$ SM-like}

\begin{figure}[h]
\begin{center}
\includegraphics[width=8cm,height=6cm]{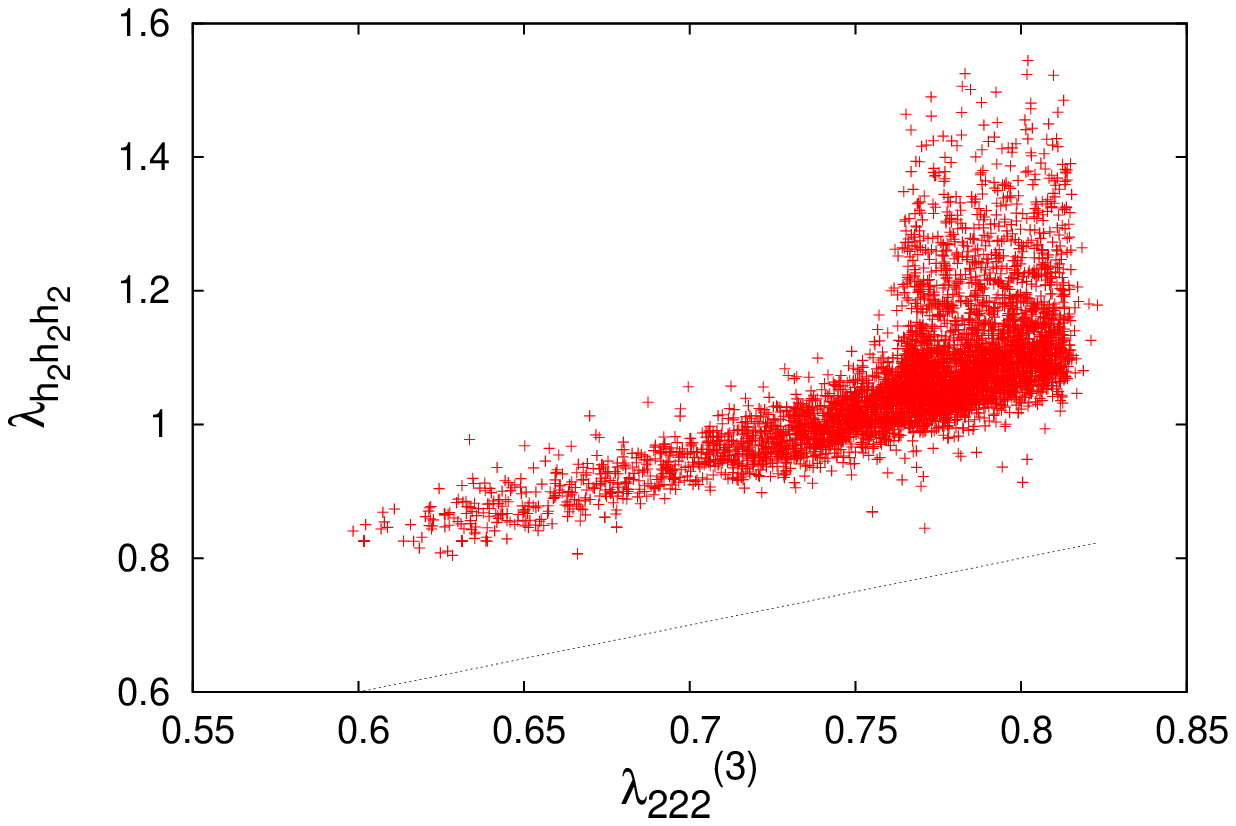}\includegraphics[width=8cm,height=6cm]{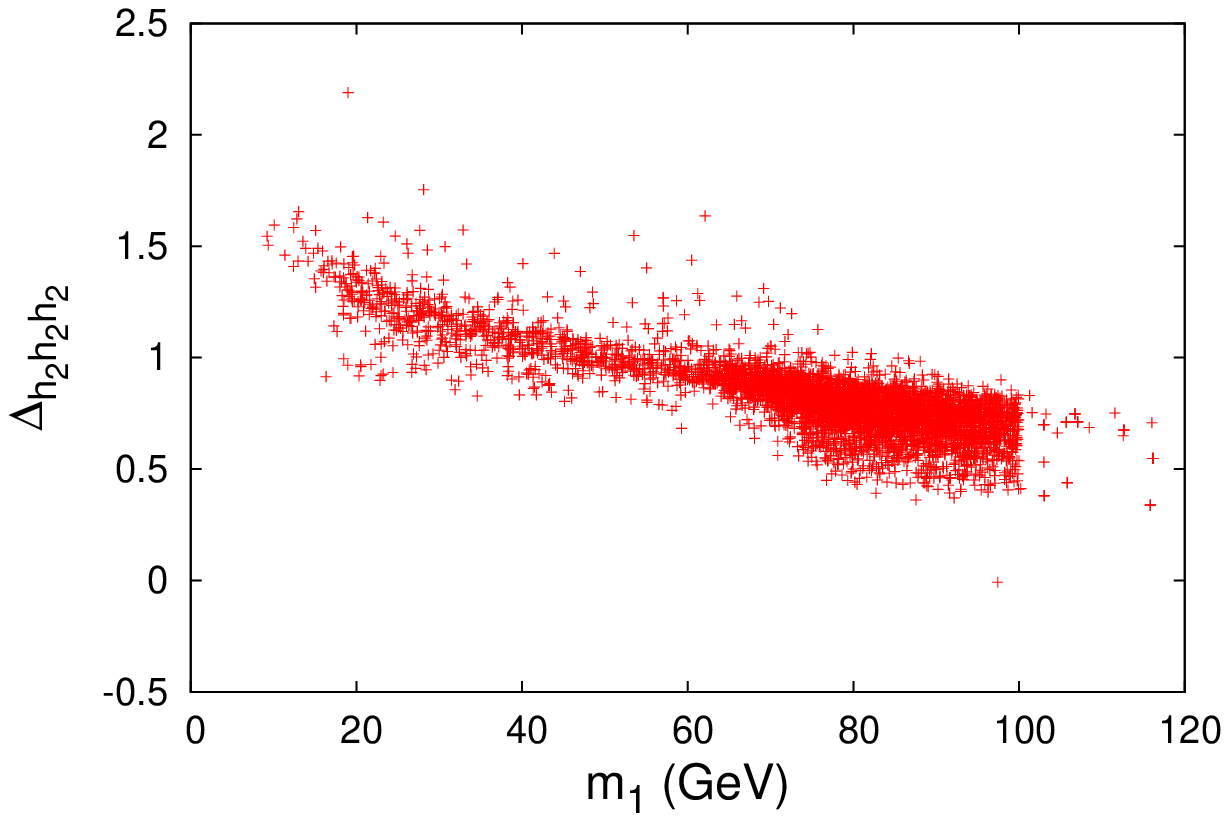}
\end{center}
\caption{\textit{Left panel: the triple Higgs couplings in absolute
values, in units of the EW vev value, versus its tree-level value
for randomly chosen sets of parameters, where the SM-like Higgs is
defined according to case B. Right panel: the relative enhancement
on the SM-like Triple Higgs coupling of $h_{2}$ with respect to the
SM case versus the mass of the light scalar $m_{1}$ for randomly
chosen sets of parameters.}
}%
\label{D2}%
\end{figure}

In the case where $h_{2}$ is SM-like, $h_{1}$ is dominated by
singlet component and according to our convention is lighter than
$h_{2}$. In this case, the relevant triple Higgs coupling is
$\lambda_{h_{2}h_{2}h_{2}}$, which corresponds to $\lambda_{hhh}$\
in the SM case. Like in case A, the other two couplings
$\lambda_{h_{1}h_{2}h_{2}}$ and $\lambda_{h_{1}h_{1}h_{2}}$ have at
least one $h_{2}$ leg which could give access to an associate
production $h_{2}h_{1}$ or double production $h_{1}h_{1}$ through
the processes:
$pp\rightarrow h_{2}^{\ast}\rightarrow h_{1}h_{2}$, $pp\rightarrow h_{2}%
^{\ast}\rightarrow h_{1}h_{1}$ at the LHC or $e^{+}e^{-}\rightarrow
Zh_{2}^{\ast}\rightarrow Zh_{1}h_{2}$, $e^{+}e^{-}\rightarrow Zh_{2}^{\ast
}\rightarrow Zh_{1}h_{1}$ at an $e^{+}e^{-}$ machine. Fig. \ref{D2}-left shows
the one-loop correction effects to the SM-like triple coupling versus its
tree-level value.

In this case, the one-loop corrections to the coupling $\lambda_{h_{2}%
h_{2}h_{2}}$\ make it larger than its corresponding tree level value. In Fig.
\ref{D2}-right, we plot the quantity $\Delta_{h_{2}h_{2}h_{2}}=(\lambda
_{h_{2}h_{2}h_{2}}-\lambda_{hhh}^{SM})/{\lambda_{hhh}^{SM}}$ as a function of
the light scalar mass $m_{1}$.

We see that in this case, the one-loop corrections to the SM-like Higgs
$h_{2}$ could enjoy large enhancement which lies between few 40\% and 100\%
for $10~\mathrm{GeV}<m_{1}<100~\mathrm{GeV}$. This effect is even amplified
and can reach 150\% and more when we cross the threshold $h_{2}\rightarrow
h_{1}h_{1}$ region. This kind of large radiative corrections have been also
reported in the framework of two Higgs doublet model \cite{Kanemura:2004mg}.

\section{Higgs Phenomenology}

In this section, we will discuss $h_{1}$ and $h_{2}$ phenomenology.

\subsection{Higgs Decays}

The partial decay widths of the two Higgs scalars $h_{1,2}$ into SM
particles such as $f\bar{f}$, $WW$ ($WW^{\ast}$) and $ZZ$
($ZZ^{\ast}$) is just the SM rate multiplied by
$\epsilon=\cos^{2}\theta,\sin^{2}\theta$, depending on whether
$h_{1}$ or $h_{2}$ is the decaying particle. This
$\epsilon$ factor apply also for loop mediated process such as $h_{i}%
\rightarrow\gamma\gamma,Z\gamma,gg$. The decay rate of $h_{2}\rightarrow
h_{1}h_{1}$\ is given by
\begin{equation}
\Gamma\left(  h_{2}\rightarrow h_{1}h_{1}\right)  =\frac{\left(  \lambda
_{112}^{(3)}\right)  ^{2}}{32\pi m_{2}}\left(  1-\frac{4m_{1}^{2}}{m_{2}^{2}%
}\right)  ^{\frac{1}{2}}\Theta\left(  m_{2}-2m_{1}\right)  ,
\end{equation}
and the light/heavy Higgs decay to DM final state $S_{0}$ is
\begin{equation}
\Gamma\left(  h_{i}\rightarrow S_{0}S_{0}\right)  =\frac{\left(  \lambda
_{00i}^{(3)}\right)  ^{2}}{32\pi m_{i}}\left(  1-\frac{4m_{0}^{2}}{m_{i}^{2}%
}\right)  ^{\frac{1}{2}}\Theta\left(  m_{i}-2m_{0}\right)  .
\end{equation}
Moreover, in this model $h_{2}$ can also decay to Triple Higgs $h_{1}$ if
kinematically allowed: $h_{2}\rightarrow h_{1}h_{1}h_{1}$ which would require
$m_{2}>3m_{1}$. This decay channel has three contributions: quartic term
$h_{2}h_{1}h_{1}h_{1}$, contribution mediated by off-shell $h_{1}^{\ast} $:
$h_{2}\rightarrow h_{1}h_{1}^{\ast}\rightarrow h_{1}h_{1}h_{1}$ and a
contribution mediated by off-shell $h_{2}^{\ast}$: $h_{2}\rightarrow
h_{1}h_{2}^{\ast}\rightarrow h_{1}h_{1}h_{1}$. This decay, even if it is open
could not compete with the 2 body phase space decay $h_{2}\rightarrow
h_{1}h_{1}$ due to the 3 body phase space suppression.

The reduction factor for the SM final state process $h_{1}\rightarrow X_{SM}$
is given by
\begin{align}
R_{X_{SM}}\left(  h_{1}\right)   &  =G\frac{B(h_{1}\rightarrow X_{SM})}%
{B^{SM}(h\rightarrow X_{SM})}\nonumber\\
&  =\frac{c^{4}\Gamma_{tot}^{SM}\left(  h_{1}\right)  }{c^{2}\Gamma_{tot}%
^{SM}\left(  h_{1}\right)  +\Gamma(h_{1}\rightarrow S_{0}S_{0})},
\end{align}
with the G-factor is given by
\begin{equation}
G=\frac{\sigma(gg\rightarrow h_{1})}{\sigma^{SM}(gg\rightarrow h)}=c^{2}.
\end{equation}
The reduction factor for $h_{2}\rightarrow X_{SM}$ is
\begin{equation}
R_{X_{SM}}\left(  h_{2}\right)  =\frac{s^{4}\Gamma_{tot}^{SM}\left(
h_{2}\right)  }{s^{2}\Gamma_{tot}^{SM}\left(  h_{2}\right)  +\Gamma
(h_{2}\rightarrow X_{NSM})},
\end{equation}
where $X_{NSM}$ denotes all the non SM final states such as
$h_{1}h_{1}$, $h_{1}h_{1}h_{1}$, $S_{0}S_{0}$ or $h_{1}S_{0}S_{0}$.
For case B, due
to the fact that $R_{X_{SM}}\left(  h_{2}\right)  $ is proportional to $s^{2}%
$, all values of $s^{2}<0.1$ will be in perfect agreement with ATLAS
and CMS data.

In the following plots, we will show our numerical results illustrating
different physical quantities for the case A previously introduced where
$h_{1}$ is the SM-like Higgs.

\begin{figure}[h]
\begin{center}
\includegraphics[width=7.6cm,height=5.8cm]{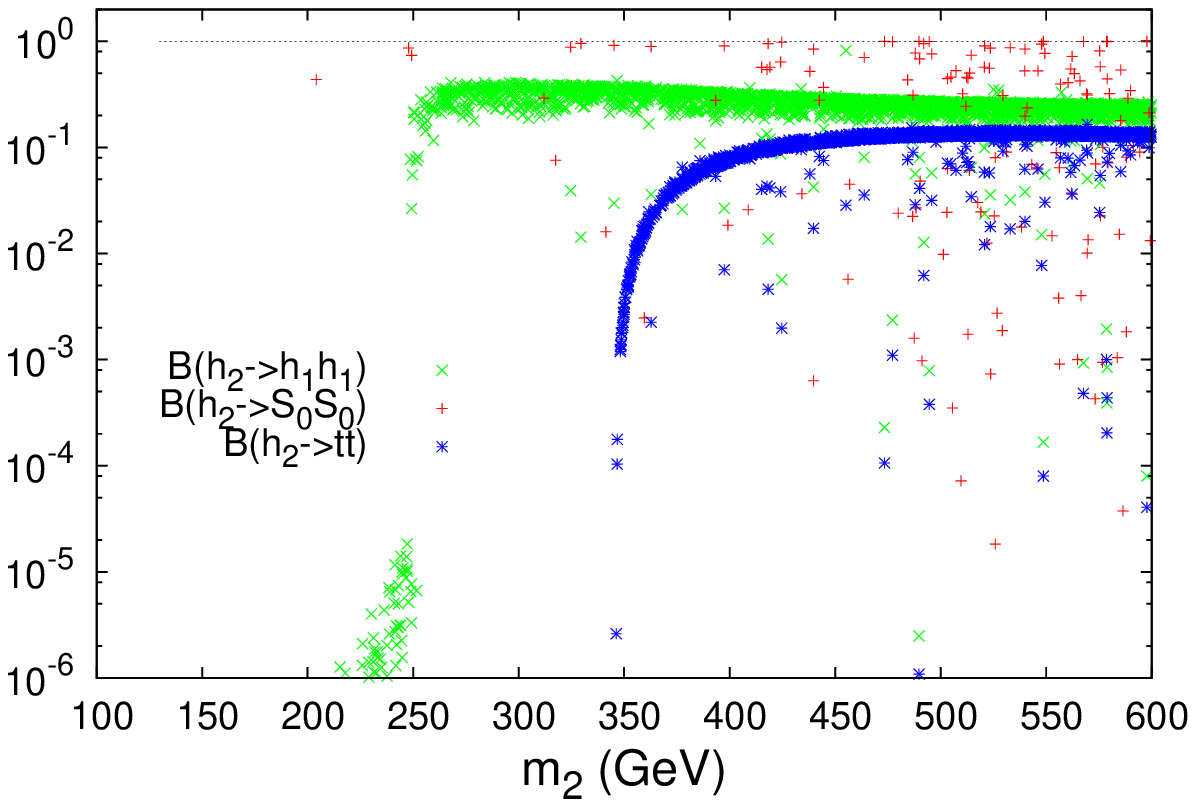}~\includegraphics[width=7.6cm,height=5.8cm]{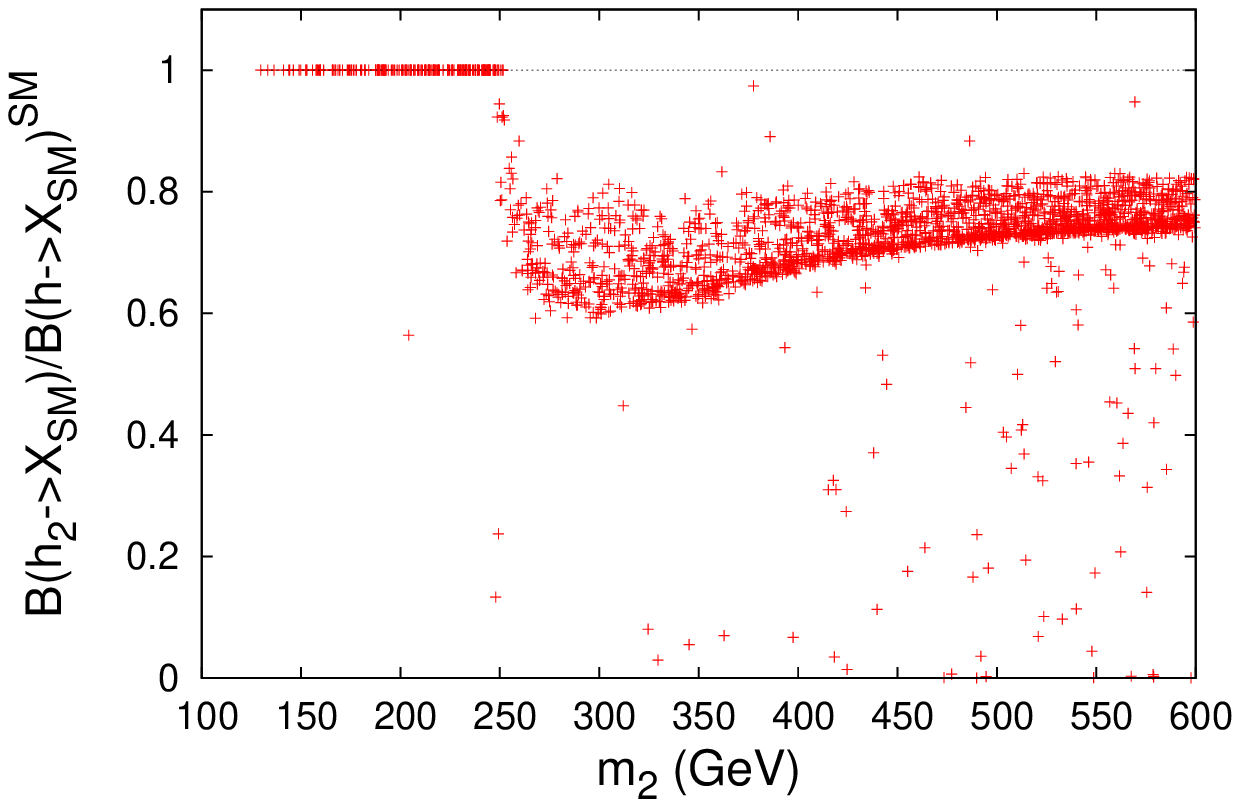}
\end{center}
\caption{\textit{Left panel: the branching ratios of the decay
channels
}$h_{2}\rightarrow t\bar{t},~h_{2}\rightarrow h_{1}h_{1}$\textit{ and }%
$h_{2}\rightarrow S_{0}S_{0}$\textit{ versus its mass. Right panel: the
branching ratio of the heavy Higgs to the SM particles final states versus the
heavy Higgs mass; scaled by the same quantities evaluated in the SM.}}%
\label{A1}%
\end{figure}

In Fig.~\ref{A1}, we show the branching ratios of $h_{2}$-decay to SM (right)
and non-SM (left) final states. In Fig.~\ref{A1}(left), we illustrate the
branching ratio of the heavy Higgs $h_{2}$ into $t\bar{t}$, $S_{0}S_{0}$ and
$h_{1}h_{1}$ as a function of $m_{2}$. It is clear that for $m_{2}%
\approx125-150$ \textrm{GeV}, $h_{2}$ will decay dominantly to SM particles
such as $b\bar{b}$, $WW^{\ast}$ and $ZZ^{\ast}$, if the decay $h_{2}%
\rightarrow S_{0}S_{0}$ is kinematically forbidden. Once $h_{2}\rightarrow
S_{0}S_{0}$ is open, it dominates all the other decays. For the range
$m_{2}\approx150-250$ \textrm{GeV}, we can see the opening of the three body
phase space channel $h_{2}\rightarrow h_{1}h_{1}^{\ast}\rightarrow h_{1}%
f\bar{f}$ which is rather small (less than $10^{-4}$). However, once
$m_{2}\geq250$ \textrm{GeV} the on-shell decay $h_{2}\rightarrow h_{1}h_{1}$
is open and compete with $h_{2}\rightarrow S_{0}S_{0}$. As one can see, the
channels $h_{2}\rightarrow h_{1}h_{1}$ and $h_{2}\rightarrow t\bar{t}$ can
reach 40\% and 10\% branching ratio respectively.

As a summary, if the invisible channel $h_{2}\rightarrow S_{0}S_{0}$ does not
dominate, one can say that:

\begin{itemize}
\item[(1)] for $m_{\mathbf{2}}<250$ \textrm{GeV}, $h_{\mathbf{2}}$ Higgs
decays similar to the SM case,

\item[(2)] for 250 \textrm{GeV} $<m_{\mathbf{2}}<400$ \textrm{GeV}, it decays
similar to the SM by 60\% and to $h_{\mathbf{1}}h_{\mathbf{1}}$ by 40 \%;

\item[(3)] for $m_{\mathbf{2}}>400$ \textrm{GeV}, $B(h_{2}\rightarrow
h_{1}h_{1})$ becomes 30\% and $B(h_{2}\rightarrow t\bar{t})$ becomes important
as 10\%.
\end{itemize}

\begin{figure}[h]
\begin{center}
\includegraphics[width=7.6cm,height=5.8cm]{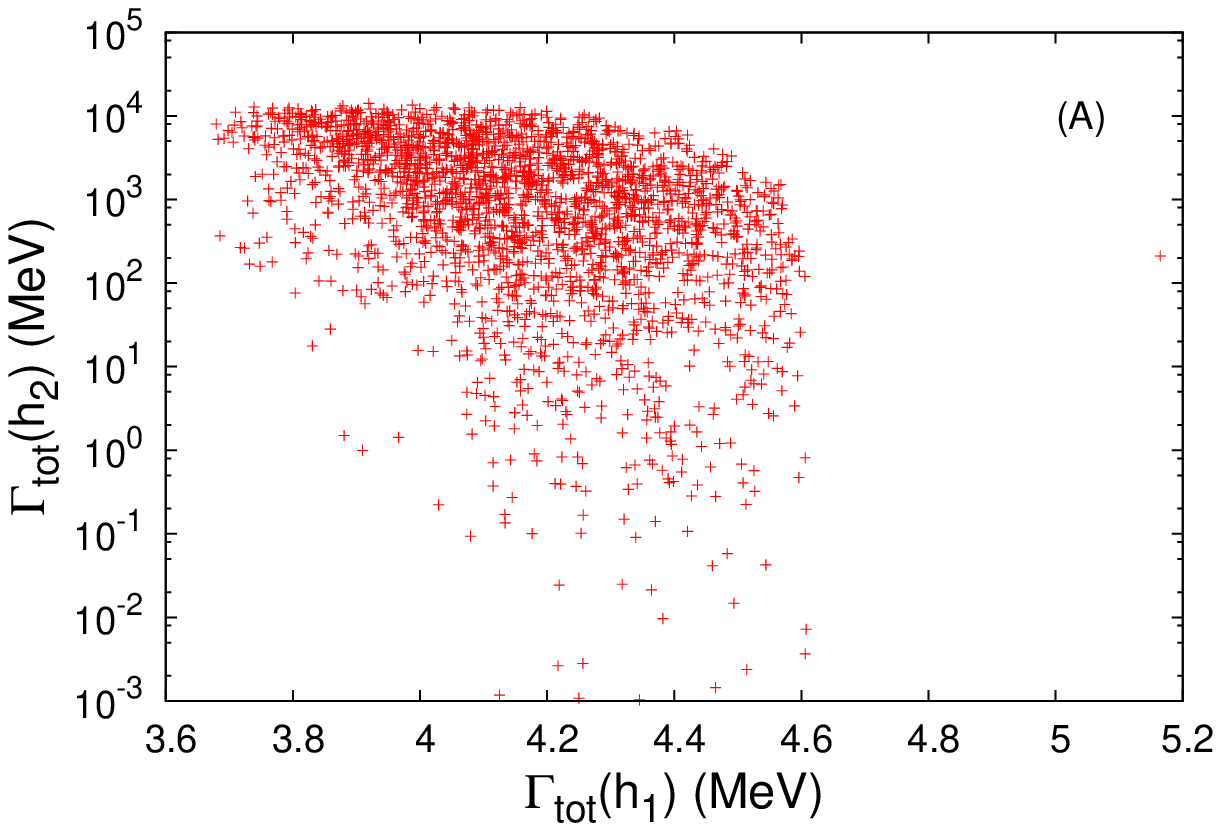}~\includegraphics[width=7.6cm,height=5.8cm]{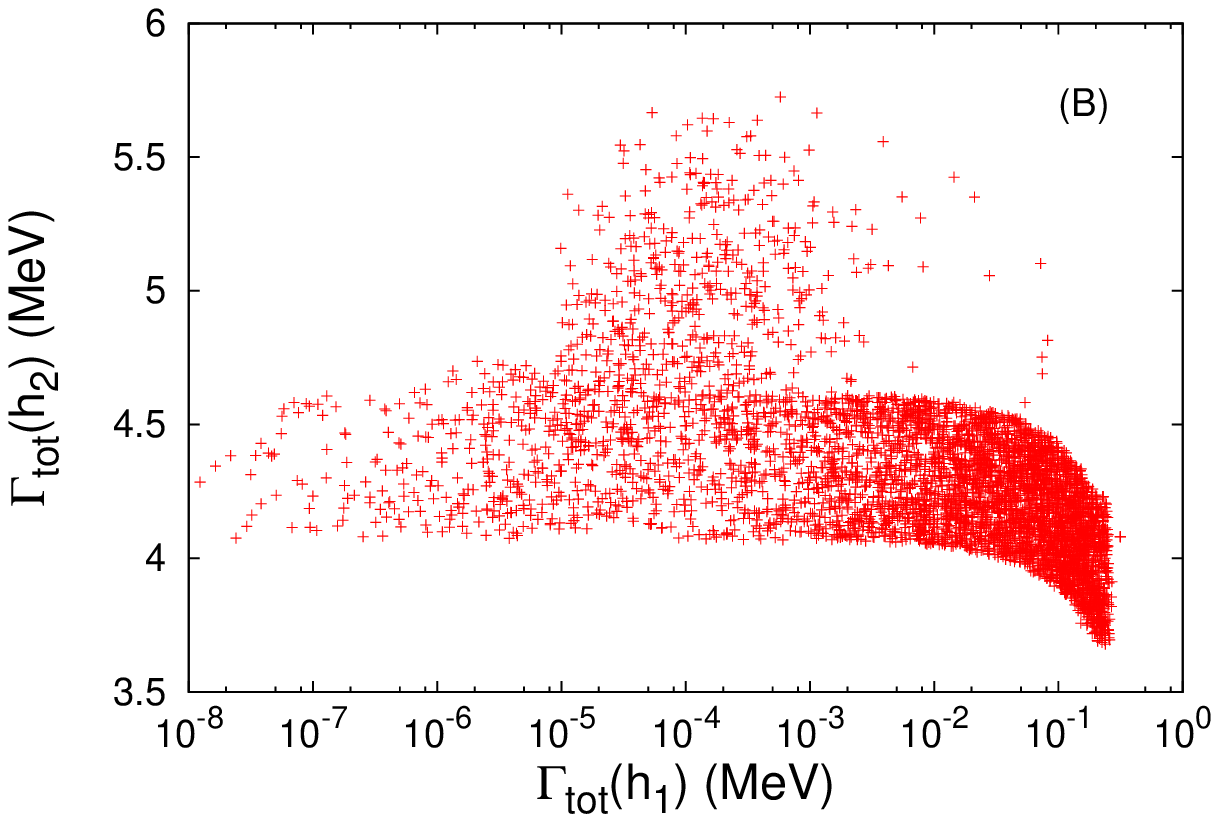}
\end{center}
\caption{\textit{The total decay width of the SM-like Higgs versus
the non
SM-like Higgs in both cases A (left) and B (right).}}%
\label{GG}%
\end{figure}

At the end, we give the total decay width for the two CP-even scalars in both
cases in Fig. \ref{GG}. It is well known that the SM Higgs with a mass of 125
\textrm{GeV} has a very narrow width which is $\Gamma_{h}\approx4$
\textrm{MeV}.

In case A where $h_{1}$ is the SM-like, the total width of $h_{1}$
is in the range 3.7-4.6 \textrm{MeV} while the total width of
$h_{2}$ can be located between $10^{-3}$ and $10^{4}$ \textrm{MeV}.
A very narrow width of $h_{2}$ means that Higgs to Higgs decays of
$h_{2}$ such that $h_{2}\rightarrow h_{1}h_{1}$ and
$h_{2}\rightarrow S_{0}S_{0}$ are closed and only $h_{2}$ decays to
SM particles are open which are suppressed because $h_{2}$ is
dominated by singlet.

In case B where $h_{2}$ is the SM-like, its total width is very
narrow (3.5-4.6 \textrm{MeV}) if $h_{2}\rightarrow S_{0}S_{0}$ and
$h_{2}\rightarrow h_{1}h_{1}$ are closed. Once these two channels
are open, the total width of $h_{2}$ grows up to 5.7 \textrm{MeV}.
The total width of $h_{1}$ which is dominated by singlet is rather
small, less than 0.2 \textrm{MeV}.

\begin{figure}[h]
\begin{center}
\includegraphics[width=7.6cm,height=5.8cm]{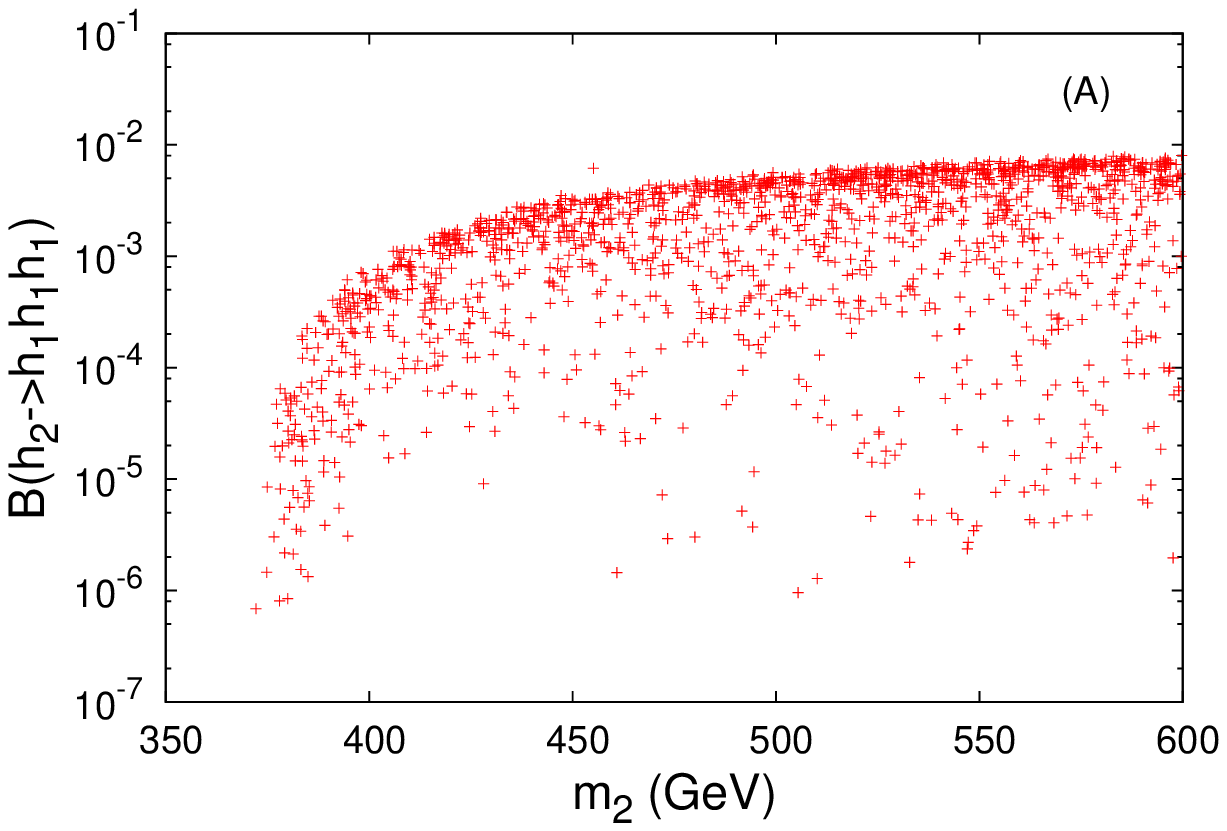}~\includegraphics[width=7.6cm,height=5.8cm]{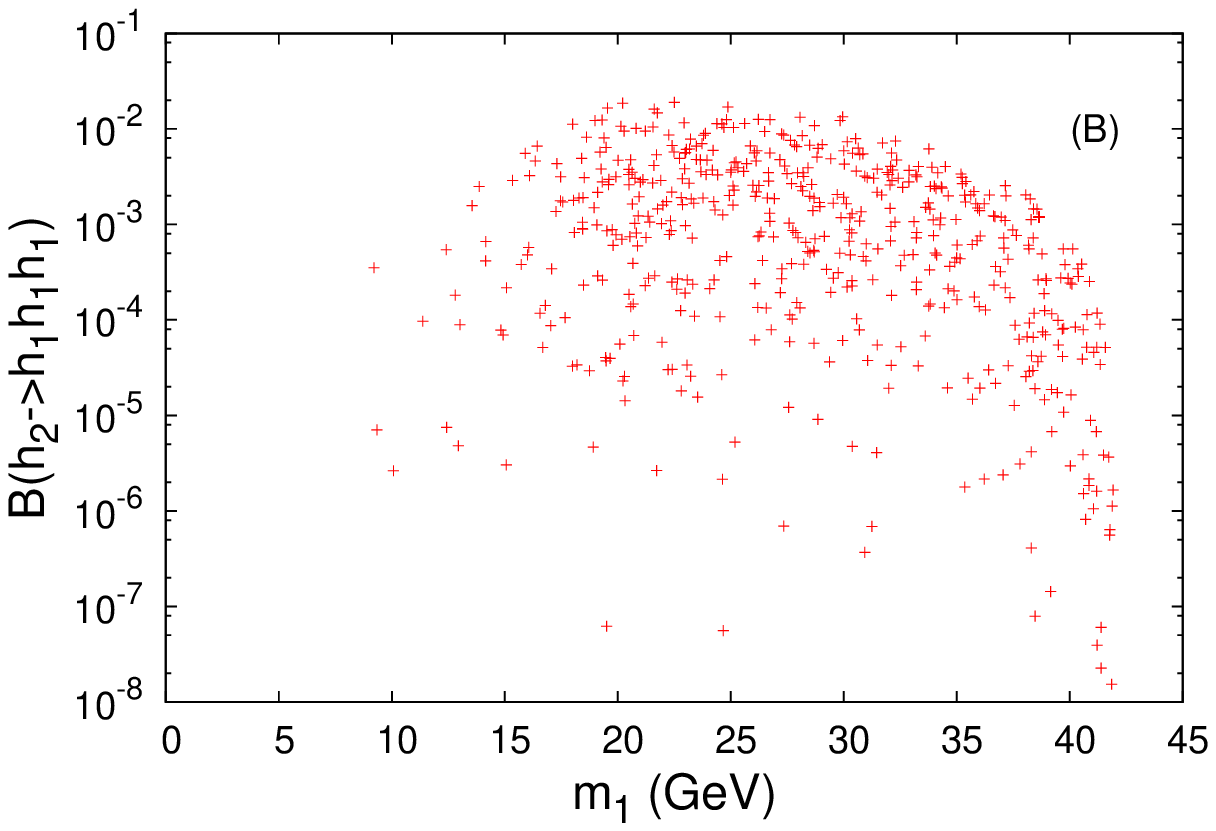}
\end{center}
\caption{\textit{The branching ratio B($h_{2}\rightarrow h_{1} h_{1} h_{1}$)
versus the non SM-like Higgs mass for both cases A (up-left) and B
(up-right).}}%
\label{h2h1h1h1}%
\end{figure}

For some benchmarks in both cases, the decay $h_{2}\rightarrow h_{1}h_{1}%
h_{1}$ is kinematically possible, and it is important to estimate how large is
this branching ratio. In Fig. \ref{h2h1h1h1}, we show $B(h_{2}\rightarrow
h_{1}h_{1}h_{1})$ versus $m_{2}$ ($m_{1}$) for case A (B).

It is clear that this branching ratio is in the order of $\mathcal{O}
(10^{-2})$ and below. We stress here that in case where $h_{2}$ is the SM-like
Higgs boson, which has quite substantial cross section, it may be possible to
measure such 3-body phase space decay with a branching ratio of the order
$10^{-2}$.

\begin{figure}[h]
\begin{center}
\includegraphics[width=7.6cm,height=5.8cm]{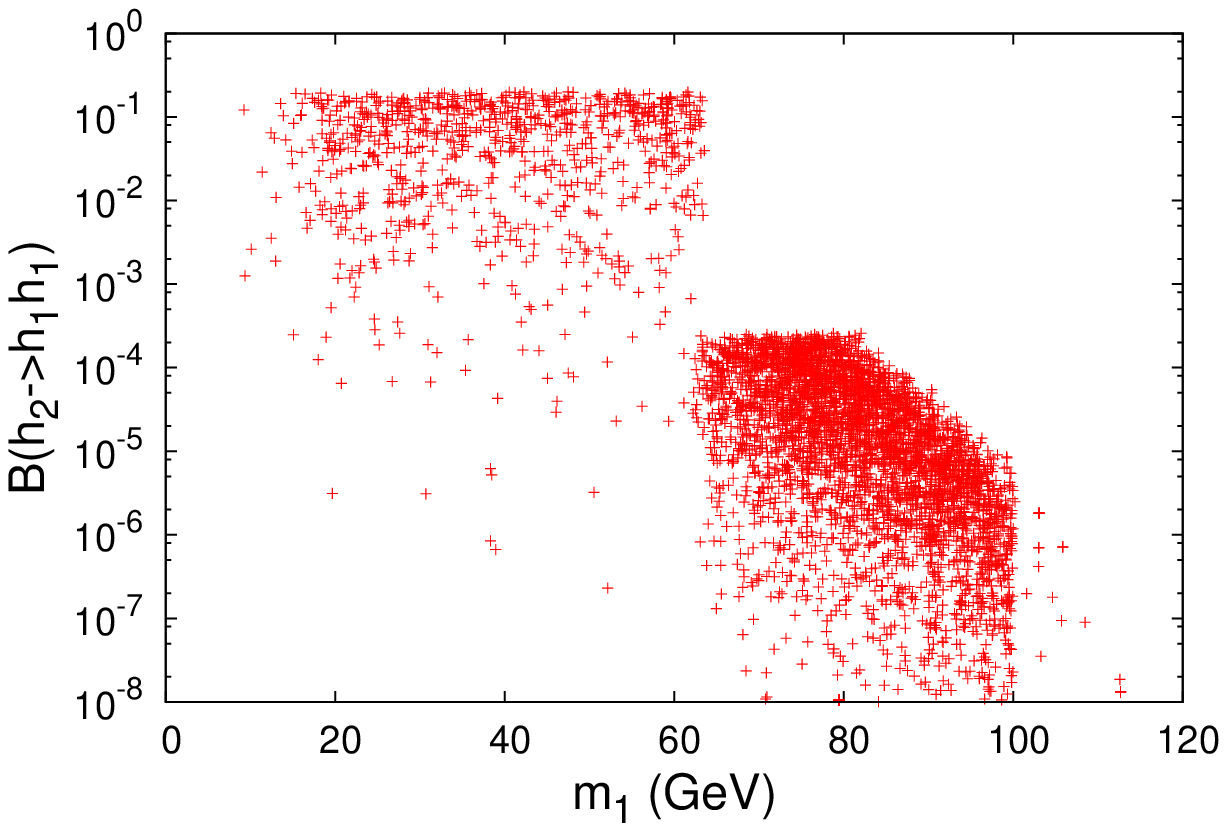}
\end{center}
\caption{\textit{The branching ratio B($h_{2}\rightarrow h_{1} h_{1}$) versus
the non SM-like Higgs mass for case B.}}%
\label{h2h1h1}%
\end{figure}

\begin{figure}[h]
\begin{center}
\includegraphics[width=7.6cm,height=5.8cm]{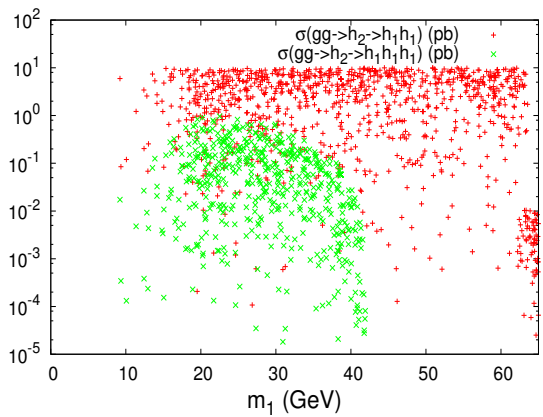}
\end{center}
\caption{\textit{The resonant production cross section for double and triple
singlet $h_{1}$ as a function of $m_{1}$ for case B. The red points are for
$\sigma(gg\to h_{2})\times B(h_{2}\to h_{1}h_{1})$, and the green ones are for
$\sigma(gg\to h_{2})\times B(h_{2}\to h_{1}h_{1}h_{1})$, all cross sections
are in pb.}}%
\label{SGG}%
\end{figure}

For case B, we show in Fig. \ref{h2h1h1} the branching ratio for
$h_{2}\rightarrow h_{1}h_{1}$ (including $h_{2}\rightarrow h_{1}h_{1}^{\ast}$)
versus the light Higgs mass; and the resonant production cross section of both
$gg\rightarrow h_{2}\rightarrow h_{1}h_{1}$ and $gg\rightarrow h_{2}%
\rightarrow h_{1}h_{1}h_{1}$ versus the light Higgs mass is shown in Fig.
\ref{SGG}.

\subsection{Higgs Production}

Same as in the SM, at the LHC the dominant production cross section for the SM
like Higgs $h_{1}$ or $h_{2}$ would be dominated by gluon fusion process which
is mediated by the top loops. The cross section rate for a single Higgs
production will be simply modified by the mixing angle $c^{2}$ or $s^{2}$
depending on $h_{1}$ or $h_{2}$ production:
\begin{align}
&  \sigma(gg\rightarrow h_{1})=c^{2}\times\sigma(gg\rightarrow h_{SM}%
)\quad,\quad\sigma(gg\rightarrow h_{2})=s^{2}\times\sigma(gg\rightarrow
h_{SM})\nonumber\\
&  \sigma(pp\rightarrow Vh_{1})=c^{2}\times\sigma(pp\rightarrow Vh_{SM}%
)\quad,\quad\sigma(pp\rightarrow Vh_{2})=s^{2}\times\sigma(pp\rightarrow
Vh_{SM})
\end{align}
It is clear that in case A where $h_{1}$ is the SM-like and dominated by
doublet component $c\approx1$ and $h_{2}$ is dominated by singlet component.
In this case, the cross section $\sigma(gg\rightarrow h_{1})$ (or
$\sigma(pp\rightarrow Vh_{1})$) will be typically close to SM one while
$\sigma(gg\rightarrow h_{2})$ (or$\ \sigma(pp\rightarrow Vh_{2})$) will be
suppressed by $s^{2}$ which is rather small in this case. Same thing apply for
the case B.

For the double Higgs production which is a good probe for Triple Higgs
self-coupling, we will evaluate $gg\rightarrow h_{i}h_{j}$ for the LHC and
$e^{+}e^{-}\rightarrow Zh_{i}h_{j}$ for the ILC in some benchmark scenarios
which are given in Table-\ref{T1} and Table-\ref{T2}. We remind here that in
the SM, the double Higgs production at the LHC $gg\rightarrow hh$ proceeds at
one-loop level trough vertex and boxes contributions (top exchange) which
interfere destructively in the total cross section. In the two singlets model
under consideration, the vertex contributions can be mediated by the 2 Higgs
scalars $h_{1,2}$: $gg\rightarrow h_{1,2}^{\ast}\rightarrow h_{i}h_{j}$ which
could give some resonant effects from $h_{2}\rightarrow h_{1}h_{1}$.

\subsubsection{Resonant Production of the SM-like Higgs}

In case A, where $h_{1}$ is the SM-like Higgs and $h_{2}$ is dominated by
singlet component. In this case the processes $gg\rightarrow h_{1}^{\ast
},h_{2}^{\ast}\rightarrow h_{1}h_{1}$ or $e^{+}e^{-}\rightarrow Zh_{1}^{\ast
},Zh_{2}^{\ast}\rightarrow Zh_{1}h_{1}$ could enjoy the resonance production
of $h_{1}h_{1}$ through the decay $h_{2}\rightarrow h_{1}h_{1}$ which could
have a branching ratio up to 20\% if open (see Table-\ref{T1}). Similar
behavior had been noticed for general two Higgs doublet model
\cite{Arhrib:2009hc}, portal model \cite{Dolan:2012ac} and next minimal
supersymmetric standard model \cite{Ellwanger:2013ova}.

As one can see from Table-\ref{T1}, the production cross section of
$h_{1}h_{1}$ could be substantial due to resonant contribution from
$h_{2}\rightarrow h_{1}h_{1}$. In the narrow width approximation of
$h_{2}$, the pair production of $h_{1}$ could be approximated by
$\sigma(pp\rightarrow h_{2})\times B(h_{2}\rightarrow h_{1}h_{1})$.
This product could be sizable if the singlet component $s$ of
$h_{2}$ is not very small and $B(h_{2}\rightarrow h_{1}h_{1})$ not
very suppressed.

\begin{table}[t]
\begin{adjustbox}{max width=\textwidth}
\begin{tabular}[c]{|c|c|c|c|c|c|c|c|}\hline
($\sin\theta$ , $m_{1}$ , $m_{2}$ , $m_{0}$) & $\lambda_{111}$ &
$\lambda_{112}$ & $\lambda_{122}$ & $B(h_{2}\rightarrow h_{1}h_{1})$
&
$\sigma^{LHC}(h_{1}h_{1})$ & $\sigma^{LHC}(h_{1}h_{2})$ & $\sigma^{ILC}%
(Zh_{1}h_{1})$\\\hline ($0.041$ , $125.9$ , $252.2$ , $415.2$) &
$195.1$ & $-15.6$ & $18.2$ & 0.1039 &
$24.3$ & $0.045$ & $0.157$\\
  & $143.8$ & $-17.7$ & $17.0$ & 0.0819 & $30.3$ & $0.041$ &
$0.134$\\\hline ($-0.16$ , $124.4$ , $249.2$ , $658.9$) & $183.2$ &
$66.2$ & $-49.3$ & 0.0349 &
$28.9$ & $0.77$ & $0.185$\\
  & $134.1$ & $76.3$ & $-44.9$ & $0.0265$ & $35.6$ & $0.65$ &
$0.175$\\\hline ($0.243$ , $124.1$ , $247.8$ , $639.7$) & $174.4$ &
$-79.8$ & $75.3$ &
7.00$\times10^{-4}$ & $22.37$ & $1.40$ & $0.167$\\
  & $129.2$ & $-92.7$ & $75.1$ & $5.18\times10^{-4}$ & $26.6$ & $1.22$
& $0.155$\\\hline ($-.23$ , $125.6$ , $550.4$ , $668.4$) & $178.6$ &
$311.5$ & $-86.5$ & 0.3057 &
$105.5$ & $0.16$ & $0.11$\\
  & $160$ & $351$ & $-45$ & 0.2573 & $115$ & $0.17$ & $0.103$\\\hline
($0.262$ , $124.3$ , $450.6$ , $802.3$) & $171.9$ & $-205.4$ &
$319.2$ &
0.1942 & $224.8$ & $0.47$ & $0.102$\\
  & $140$ & $-198$ & $311$ & $0.2065$ & $222.6$ & $0.46$ &
$0.09$\\\hline ($-.26$ , $124.1$ , $295.5$ , $920.3$) & $169$ &
$140$ & $-105$ & 0.4890 & $387$
& $1.5$ & $2.4$\\
  & $131$ & $165$ & $-80$ & $0.4074$ & $407$ & $1.18$ & $2.05$\\\hline
($-.31$ , $125.5$ , $406.1$ , $662$ ) & $165$ & $258$ & $-103$ &
0.3855 & $478$ &
$0.85$ & $0.19$\\
  & $149$ & $297$ & $-65$ & $0.3215$ & $511$ & $0.78$ & $0.17$\\\hline
\end{tabular}
\end{adjustbox}
\caption{\textit{Benchmarks scenario for case A, all masses,
couplings $\lambda_{ijk}$ are in GeV. The LHC energy at 14 TeV and
the ILC at 500 GeV, all cross sections are in fb. In the SM,
$\sigma^{LHC}(pp\rightarrow hh)=25.4$ fb at 14 TeV and
$\sigma^{ILC}(e^{+}e^{-}\rightarrow Zhh)=0.14$ fb at 500 GeV for
$m_{h}=125$ GeV. The first value for triple couplings, branching
ratio $B(h_2\to h_1h_1)$ and cross sections corresponds to the
Leading Order (LO) while the second one corresponds to
 improved LO by taking the triple coupling at one-loop level.}}
\label{T1}%
\end{table}

In Table-\ref{T1}, the benchmarks in lines 4, 5 and 7 correspond to
the case where the decay $h_{2}\rightarrow h_{1}h_{1}h_{1}$ is
kinematically possible; however its branching ratio is
$1.022\times10^{-3}$, $2.178\times10^{-5}$ and $1.352\times10^{-4}$,
for these benchmarks respectively. For case A, the branching ratio
$B(h_{2}\rightarrow h_{1}h_{1})$, the coupling
$\lambda_{h_{1}h_{1}h_{2}}$ and the cross sections
$\sigma(pp\rightarrow h_{1}h_{1})$ and $\sigma(e^{-}e^{+}\rightarrow
Zh_{1}h_{1})$ could receive corrections up to 20\%, 16\%, 26\% and
14\%, respectively.

\begin{table}[t]
\centering
\begin{adjustbox}{max width=\textwidth}
\begin{tabular}[c]{|c|c|c|c|c|c|c|c|}\hline
($\sin\theta$ , $m_{1}$ , $m_{2}$ , $m_{0}$) & $\lambda_{112}$ &
$\lambda_{122}$ & $\lambda_{222}$ & $B(h_{2}\rightarrow h_{1}h_{1})$
& $\sigma^{LHC}(h_{2}h_{2})$ & $\sigma^{LHC}(h_{1}h_{2})$ &
$\sigma^{ILC} (Zh_{2}h_{2})$\\\hline ($0.98$ , $63$ , $126.2$ ,
$281.5$) & $16$ & $34$ & $-183.5$ & $0.1171$ & $3555$
& $22.8$ & $35.19$\\
  & $13$ & $17.5$ & $-262.3$ & $0.1696$ & $2421$ & $16.2$ &
$30.8$\\\hline ($0.999$ , $54.8$ , $124$ , $584.7$) & $4.8$ & $3.46$
& $-189.38$ & 0.1842 &
$3715.86$ & $25.069$ & $39.67$\\
  & $4.9$ & $1.2$ & $-277$ & $0.1718$ & $3992$ & $17$ &
$28$\\\hline ($-0.976$ , $85.5$ , $126.9$ , $263.4$) & $68.4$ & $16$
& $189.7$ &
5.29$\times10^{-3}$ & $0.92$ & $21.95$ & $0.0061$\\
  & $73.2$ & $3.47$ & $252.3$ & $4.61\times10^{-3}$ & $1.34$ & $15.8$ &
$0.008$\\\hline ($-0.966$ , $87.2$ , $125.9$ , $480.3$) & $79.8$ &
$15.1$ & $184.2$ &
4.92$\times10^{-3}$ & $0.868$ & $21.33$ & $0.008$\\
  & $86.7$ & $67.2$ & $244.9$ & $4.16\times10^{-3}$ & $1.39$ & $16.4$ &
$0.01$\\\hline ($0.967$ , $94.5$ , $126.1$ , $191.4$) & $56.9$ &
$60.9$ & $-171.2$ &
5.06$\times10^{-4}$ & $0.57$ & $21.66$ & $0.0005$\\
  & $56.5$ & $43$ & $-226.2$ & $5.13\times10^{-4}$ & $0.63$ &
$17.2$ & $0.0005$\\\hline ($0.977$ , $82$ , $124.5$ , $862$ )&
$52.5$ & $48.016$ & $-174.509$ &
4.36$\times10^{-3}$ & $1.308$ & $22.95$ & $0.0017$\\
  & $52.6$ & $33.2$ & $-231.8$ & $4.35\times10^{-3}$ & $1.43$ & $18$ &
$0.002$\\\hline
\end{tabular}
\end{adjustbox}
\caption{\textit{Benchmarks scenario for case B, all masses,
couplings $\lambda_{ijk}$ are in GeV. The LHC energy at 14 TeV and
the ILC at 500 GeV, all cross sections are in fb. The first value
for triple couplings, branching ratio $B(h_2\to h_1h_1)$ and cross
sections corresponds to the leading order (LO) while the second one
corresponds to
 improved LO by taking the triple coupling at one-loop level.}}
\label{T2}%
\end{table}

In case B, since $h_2$ is SM-like, the production $pp\rightarrow
h_{2}h_{2}$, will be roughly similar to SM, because in this case
$h_{1}$ is lighter than $h_{2}$ and then $pp\rightarrow h_{2}h_{2}$
can not benefit from the resonant production of $h_{1}$ to a pair of
$h_{2}$. This can be seen in Table-\ref{T2}. We stress here that by
taking the values of the couplings $\lambda_{122}$ and
$\lambda_{222}$ at one-loop level reduces slightly the cross
sections as can be seen from Table-\ref{T2}. In the first benchmark
of Table-\ref{T2}, improving the coupling $\lambda_{112}$ by the one
loop corrections can modify the Branching ratio $B(h_2\to h_1h_1)$
 and the cross section $gg\to h_1h1$ up
to 45\% and 32\% respectively.

\subsubsection{Singlet Scalars Production}

As we have seen previously for case B, the decay $h_{2}\rightarrow h_{1}h_{1}
$ could be open and its branching ratio could reach 20\%. Using the fact that
a SM-Higgs $h_{2}$ with 125 \textrm{GeV} will be copiously produced at 14
\textrm{TeV} LHC: $\sigma(pp\rightarrow h_{2})\approx50$ \textrm{pb}, one can
have access to the following production for two singlet scalars:
\begin{align}
\sigma(pp\rightarrow h_{1}h_{1}) & \approx\sigma(pp\rightarrow
h_{2})\times
B(h_{2}\rightarrow h_{1}h_{1})\nonumber\\
& \approx s^{2}\times\sigma(pp\rightarrow h_{SM})\times
B(h_{2}\rightarrow
h_{1}h_{1})\nonumber\\
& \approx9\left[ \frac{s^{2}}{0.9}\right] \left[
\frac{B(h_{2}\rightarrow h_{1}h_{1})}{0.2}\right]
(\mathrm{pb})\text{ for }m_{2}=125\ \mathrm{GeV},
\end{align}
which is rather substantial if $B(h_{2}\rightarrow h_{1}h_{1})$ is not
suppressed. As it is illustrated in Fig. \ref{SGG} (red points), the
production cross section for double singlet $h_{1}$ could be substantial for
large area of parameter space and can reach 10 \textrm{pb} which would lead to
a visible signal if this scenario is realized.

Same estimate for triple Higgs production at the LHC gives:
\begin{align}
\sigma(pp\rightarrow h_{1}h_{1}h_{1}) & \approx\sigma(pp\rightarrow
h_{2})\times B(h_{2}\rightarrow h_{1}h_{1}h_{1})\nonumber\\
& \approx s^{2}\times\sigma(pp\rightarrow h_{SM})\times
B(h_{2}\rightarrow
h_{1}h_{1}h_{1})\nonumber\\
& \approx0.1\left[ \frac{s^{2}}{0.9}\right] \left[ \frac{B(h_{2}%
\rightarrow h_{1}h_{1}h_{1})}{10^{-2}}\right] (\mathrm{pb})\text{ for }%
m_{2}=125~\mathrm{GeV},
\end{align}
which is rather large compared to the Drell-Yann cross section for
$2\rightarrow2$ processes. In Fig.~\ref{SGG} (green points), we illustrate the
production cross section for triple singlet $h_{1}$ coming mainly from
$\sigma(gg\rightarrow h_{2})\times B(h_{2}\rightarrow h_{1}h_{1}h_{1})$. As it
can be seen it turns out that this production channel could give a cross
section up to 1 \textrm{pb} if the singlet $h_{1}$ is in the range 20-30
\textrm{GeV}. At the 14 TeV LHC run with 100 \textrm{fb}$^{-1}$ luminosity,
$1$ \textrm{pb} cross section can leads to $10^{5}$ raw events of $6b$ or
$4b2\tau$ or $2b4\tau$ or $6\tau$ without cuts.

Similarly, at the ILC we can have access to a pair of singlet scalars by
producing first the SM Higgs $h_{2}$ which can decay with sizable branching
ratio to a pair of singlet scalars. In the narrow width approximation of
$h_{2}$, we have:
\begin{align}
\sigma(e^{+}e^{-}\rightarrow Zh_{1}h_{1}) & \approx\sigma(e^{+}%
e^{-}\rightarrow Zh_{2})\times B(h_{2}\rightarrow h_{1}h_{1})\nonumber\\
& \approx s^{2}\times\sigma(e^{+}e^{-}\rightarrow Zh_{SM})\times
B(h_{2}\rightarrow h_{1}h_{1})\nonumber\\
& \approx21.5\left[ \frac{s^{2}}{0.9}\right] \left[ \frac{B(h_{2}%
\rightarrow h_{1}h_{1})}{0.2}\right] (fb)\text{\quad for\quad}\sqrt
{s}=250~\mathrm{GeV}\nonumber\\
& \approx5.1\left[ \frac{s^{2}}{0.9}\right] \left[ \frac{B(h_{2}%
\rightarrow h_{1}h_{1})}{0.2}\right] (fb)\text{ for\quad}\sqrt{s}%
=500~\mathrm{GeV.}%
\end{align}
It is obvious, that at the ILC the cross section is more important near
threshold production of $Zh_{2}$ which is close to 250 \textrm{GeV}. Since the
process $e^{+}e^{-}\rightarrow Zh_{2}$ is mediated by s-channel Z exchange,
the cross section is slightly suppressed for higher center of mass energy
$\geq500$ \textrm{GeV}.

To have an idea about the order of magnitude of these cross sections
both at the LHC-14 TeV and the ILC we give some numerical results in
Table-\ref{T1} for case A and Table-\ref{T2} for case B. It is clear
that both at LHC and ILC the double Higgs production can be larger
or smaller than the corresponding SM one. In the cross sections for
hadron collider, we include a K-factor $K=2$ \cite{Dawson:1998py}.
In case A, one can see that the cross section of a pair production
of SM-like Higgs could exceed in some cases 100 fb, which would give
more than $10^{4}$ raw events for an integrated LHC luminosity of
100 fb$^{-1}$ giving rise to $b\bar{b}b\bar{b}$ and
$b\bar{b}\tau^{+}\tau^{-} $ final states with large transverse
momentum. Observation of such large Higgs pair production cross
sections would be a clear evidence for physics beyond the SM.

In case B, where $h_{1}$ is a singlet with a mass less than 125
\textrm{GeV}, it is clear from Table-\ref{T2} that pair production
of singlet scalars could be substantial and the LHC cross sections
could exceed 3 \textrm{pb}, giving more events than in the previous
case. In case B, where $h_{2}$ is the SM-like, the lighter Higgs
scalar $h_{1}$ decays to SM final states with the same branching
ratios as the SM Higgs. Then, for benchmarks where the cross section
$\sigma(pp\rightarrow h_{1}h_{1})$ is around 10 \textrm{pb}, we will
have the $b\bar{b}b\bar{b}$ final state. However, this final state
suffers from a huge QCD background. The $b\bar{b}\tau^{-}\tau^{+}$,
$b\bar{b}\gamma\gamma$ final states are promising one in the case of
SM Higgs pair production \cite{double0,double3}. Since in our case,
the production cross section is much higher than the production of a
Higgs pair in the SM, a possible signal extraction could be
performed with a very good efficiency. A more interesting final
state is $\tau^{-}\tau^{+}\tau^{-}\tau^{+}$, which would give same
sign dileptons if the $\tau$'s of the same electric charge decay
leptonically. All these possible final states need a full Monte
Carlo analysis which is out of the scope of the present study.

\section{Conclusion}

We have shown that the two-singlets model can accommodate a Higgs
boson with a mass in the range 125-126 \textrm{GeV} together with
the relic density and indirect detection constraints as well as all
the recent measurements from ATLAS and CMS experiments. The model
has three CP-even Higgs $h_{1,2}$, two of which are a mixture of
doublet and a singlet components, while the third one is a singlet
particle $S_{0}$ which plays the role of DM candidate. We studied
both the cases where $h_{1} $ or $h_{2}$ is the SM-like Higgs; and
investigated the effect of the extra Higgs bosons on the triple
Higgs self-couplings. We have found that in the case where $h_{1}$
is the SM-like Higgs, the Triple Higgs self-coupling
$h_{1}h_{1}h_{1}$ can receive a significant enhancement which could
be greater than 40\% for $m_{2}>600$ \textrm{GeV}. In the case where
$h_{2}$ is the SM-like Higgs, the Higgs triple self-coupling
$h_{2}h_{2}h_{2}$ receives an enhancement between 50\% and 150\%.

We have discussed that some of the Higgs pair $h_{i}h_{j}$ could be produced
either at the 14 TeV LHC with high luminosity option or at the future linear
collider where the mass and the triple coupling of the Higgs could be measured
with very good precision. We have also seen that when $h_{2}$ is the SM-like
Higgs and $h_{1}$ is singlet dominated Higgs and lighter than $h_{2}$, one can
produce either a pair of $h_{1}$ or triple $h_{1}$ through $\sigma
(pp\rightarrow h_{2})\times B(h_{2}\rightarrow h_{1}h_{1})$ or $\sigma
(pp\rightarrow h_{2})\times B(h_{2}\rightarrow h_{1}h_{1}h_{1})$ with
substantial cross section. This will constitute an important mechanism for
producing singlet scalars in this model. In the other case where $h_{1}$ is
the SM-like Higgs and $h_{2}$ is the singlet scalar, we have seen that we can
have a cross section of a Higgs pair $h_{1}$ which is more than one order of
magnitude larger than the corresponding SM one. Observation of such large
Higgs pair production would be a clear indication of physics beyond the SM.

\section*{Acknowledgements}

The work of A. Ahriche is supported by the Algerian Ministry of Higher
Education and Scientific Research under the PNR '\textit{Particle
Physics/Cosmology: the interface}', and the CNEPRU Project No.
\textit{D01720130042}. A. Arhrib would like to thank NSC-Taiwan for financial
support during his stay at Academia Scinica where part of this work has been done.

\appendix

\section{Cubic and Quartic Scalar Couplings}

The cubic and quartic terms are obtained after the symmetry breaking as
couplings between the scalar eigenstates. Here we used a notation where the
subscripts $0$, $1$ and $2$ denote $S_{0}$, $h_{1}$ and $h_{2}$ respectively.
The cubic couplings with dimension of a mass are
\begin{align}
\lambda_{001}^{(3)} & =c\lambda_{0}\upsilon+s\eta_{01}\upsilon_{1}%
,~\lambda_{002}^{(3)}=c\eta_{01}\upsilon_{1}-s\lambda_{0}\upsilon,\nonumber\\
\lambda_{111}^{(3)} & =c^{3}\lambda\upsilon+\frac{3}{2}s^{2}\lambda
_{1}(c\upsilon_{1}+s\upsilon)+s^{3}\eta_{1}\upsilon_{1},\nonumber\\
\lambda_{222}^{(3)} & =c^{3}\eta_{1}\upsilon_{1}-3cs\lambda_{1}%
(c\upsilon-s\upsilon_{1})-s^{3}\lambda\upsilon,\nonumber\\
\lambda_{112}^{(3)} & =c^{3}\lambda_{1}\upsilon_{1}+cs[c(2\lambda
_{1}-\lambda)\upsilon-s(2\lambda_{1}-\eta_{1})\upsilon_{1}]-s^{3}\lambda
_{1}\upsilon,\nonumber\\
\lambda_{122}^{(3)} &
=c^{3}\lambda_{1}\upsilon-cs[c(2\lambda_{1}-\eta
_{1})\upsilon_{1}+s(2\lambda_{1}-\lambda)\upsilon]+s^{3}\lambda_{1}%
\upsilon_{1}, \label{l3}%
\end{align}
and the quartic terms are
\begin{align}
\lambda_{1111}^{(4)} & =\lambda c^{4}+6\lambda_{1}c^{2}s^{2}+\eta_{1}%
s^{4},~\lambda_{2222}^{(4)}=\eta_{1}c^{4}+6\lambda_{1}c^{2}s^{2}+\lambda
s^{4},\nonumber\\
\lambda_{0011}^{(4)} & =\lambda_{0}c^{2}+\eta_{01}s^{2},~\lambda
_{0022}^{(4)}=\eta_{01}c^{2}+\lambda_{0}s^{2},~\lambda_{012}^{(4)}%
=cs(\eta_{01}-\lambda_{0}),\nonumber\\
\lambda_{1112}^{(4)} & =cs[(3\lambda_{1}-\lambda)c^{2}-(3\lambda_{1}%
-\eta_{1})s^{2}],~\lambda_{1122}^{(4)}=\lambda_{1}\left(
c^{2}-s^{2}\right)
^{2}-c^{2}s^{2}(2\lambda_{1}-\eta_{1}-\lambda),\nonumber\\
\lambda_{1222}^{(4)} & =cs[(\eta_{1}-3\lambda_{1})c^{2}-(\lambda
-3\lambda_{1})s^{2}]. \label{l4}%
\end{align}

\section{The Effective Triple Higgs Couplings}

The effective triple Higgs couplings can be estimated as the third derivatives
of the effective potential with respect the scalar CP-even eigenstates. For a
general form of the effective potential%
\begin{align}
V\left( \tilde{h}\right)  & =-%
{\textstyle\sum\limits_{k}}
\tfrac{\mu_{k}^{2}}{2}\tilde{h}_{k}^{2}+\tfrac{\lambda_{k}}{24}\tilde{h}%
_{k}^{4}+{\sum\limits_{i,k}}\tfrac{\omega_{ik}}{4}\tilde{h}_{i}^{2}\tilde
{h}_{k}^{2}+V^{1-l}(\tilde{h}),\label{VG}\\
V^{1-l}\left( \tilde{h}\right)  & ={\sum_{\alpha=all~fields}}%
\tfrac{n_{\alpha}m_{\alpha}^{4}(\tilde{h})}{64\pi^{2}}\left( \log
\tfrac{m_{\alpha}^{2}(\tilde{h})}{\Lambda^{2}}-c_{\alpha}\right) ,
\end{align}
with $\omega_{ik}=0$ for $k\leq i$, the effective triple Higgs couplings are
given by
\begin{equation}
\lambda_{ijk}^{(3)}=\lambda_{ijk}^{(3-tree)}+\tfrac{\partial^{3}}{\partial
h_{i}\partial h_{j}\partial h_{k}}\left[
{\sum_{\alpha}}\tfrac{n_{\alpha
}m_{\alpha}^{4}(\tilde{h})}{64\pi^{2}}\left( \log\tfrac{m_{\alpha}^{2}%
(\tilde{h})}{\Lambda^{2}}-c_{\alpha}\right) \right] , \label{L3eff}%
\end{equation}
where $\lambda_{ijk}^{(3-tree)}$\ are the\ tree-level triple couplings in
(\ref{l3}), $c_{\alpha}$ depends on the renormalization scheme; and $\tilde
{h}$ are the CP-even scalars (like $\tilde{h}$ and $\chi_{1}$ in our model)
and $h$ are the eigenstates after the symmetry breaking where%
\begin{equation}
h_{i}=u_{ik}\tilde{h}_{k},~\tilde{h}_{i}=u_{ik}^{T}h_{k}=u_{ki}h_{k},
\end{equation}
and $u_{ik}$\ are the mixing matrix elements. In order to evaluate the second
term in (\ref{L3eff}), we parameterize the field dependent masses as%
\begin{equation}
m_{\alpha}^{2}(\tilde{h})=m_{\alpha}^{2}\left[
1+\epsilon_{\alpha}\right] ,
\end{equation}
with $\epsilon$ in terms of the eigenstates $h_{i}$ or the fields
$\tilde {h}_{i}$; and can be expanded as
\begin{align}
\epsilon_{\alpha} & \simeq\eta_{\alpha,i}h_{i}+\varsigma_{\alpha,ik}%
h_{i}h_{k}+\xi_{\alpha,ikl}h_{i}h_{k}h_{l},\nonumber\\
&
\simeq\tilde{\eta}_{\alpha,i}\tilde{h}_{i}+\tilde{\varsigma}_{\alpha
,ik}\tilde{h}_{i}\tilde{h}_{k}+\tilde{\xi}_{\alpha,ikl}\tilde{h}_{i}\tilde
{h}_{k}\tilde{h}_{l},\label{ep}\\
\eta_{\alpha,i} & =u_{ki}\tilde{\eta}_{\alpha,k},~\varsigma_{\alpha
,ik}=u_{li}u_{mk}\tilde{\varsigma}_{\alpha,lm},~\xi_{\alpha,ikl}=u_{mi}%
u_{nk}u_{rl}\tilde{\xi}_{\alpha,mnr},
\end{align}
where there is a summation over the repeated indices. Then%
\begin{align}
\lambda_{ijk}^{(3)} & =\lambda_{ijk}^{(3-tree)}+\sum_{\alpha}\tfrac
{n_{\alpha}m_{\alpha}^{4}}{32\pi^{2}}\left[ \left( \xi_{\alpha,ijk}%
+\eta_{\alpha,i}\varsigma_{\alpha,jk}+\eta_{\alpha,k}\varsigma_{\alpha
,ij}+\eta_{\alpha,j}\varsigma_{\alpha,ik}\right) \log\tfrac{m_{\alpha}^{2}%
}{m_{h_{1}}^{2}}\right. \nonumber\\
& \left. +\xi_{\alpha,ijk}\left( \tfrac{1}{2}-c_{\alpha}\right)
+\left(
\eta_{\alpha,i}\varsigma_{\alpha,jk}+\eta_{\alpha,k}\varsigma_{\alpha,ij}%
+\eta_{\alpha,j}\varsigma_{\alpha,ik}\right) \left(
\tfrac{3}{2}-c_{\alpha
}\right) +\eta_{\alpha,i}\eta_{\alpha,j}\eta_{\alpha,k}\right] \nonumber\\
&
+\log\frac{m_{h_{1}}^{2}}{\Lambda^{2}}\sum_{\alpha}\tfrac{n_{\alpha
}m_{\alpha}^{4}}{32\pi^{2}}\left( \xi_{\alpha,ijk}+\eta_{\alpha,i}%
\varsigma_{\alpha,jk}+\eta_{\alpha,k}\varsigma_{\alpha,ij}+\eta_{\alpha
,j}\varsigma_{\alpha,ik}\right) , \label{Tri}%
\end{align}
where the scale dependance is isolated in the last line. This scale dependance
can be eliminated in favor of measurable quantities such as CP-even scalar
eigenmasses. In order to do so, let us take the general form of the scalar
effective potential (\ref{VG}). Then, the tadpole gives%
\begin{equation}
\mu_{k}^{2}=\tfrac{\lambda_{k}}{6}\upsilon_{k}^{2}+{\sum\limits_{l}}%
\tfrac{\omega_{kl}+\omega_{lk}}{2}\upsilon_{l}^{2}+{\sum_{\alpha}}%
\tfrac{n_{\alpha}\tilde{\eta}_{\alpha,k}m_{\alpha}^{4}}{32\pi^{2}\upsilon_{k}%
}\left( \log\tfrac{m_{\alpha}^{2}}{m_{h_{1}}^{2}}-c_{\alpha}+\frac{1}%
{2}\right) +\log\frac{m_{h_{1}}^{2}}{\Lambda^{2}}{\sum_{\alpha}}%
\tfrac{n_{\alpha}\tilde{\eta}_{\alpha,k}m_{\alpha}^{4}}{32\pi^{2}\upsilon_{k}%
}, \label{tadpole}%
\end{equation}
and the summation of all CP-even scalar masses taking into account the tadpole
conditions (\ref{tadpole}) are given by%
\begin{align}
{\sum_{k}}m_{h_{k}}^{2} & ={\sum_{k}}\tfrac{\lambda_{k}}{3}\upsilon_{k}%
^{2}+\tfrac{1}{32\pi^{2}}\sum_{\alpha,k}n_{\alpha}m_{\alpha}^{4}\left(
\tilde{\eta}_{\alpha,k}^{2}+\tilde{\varsigma}_{\alpha,kk}-\frac{\tilde{\eta
}_{\alpha,k}}{\upsilon_{k}}\right) \log\tfrac{m_{\alpha}^{2}}{m_{h_{1}}^{2}%
}\nonumber\\
& +\tfrac{1}{32\pi^{2}}\sum_{\alpha,k}n_{\alpha}m_{\alpha}^{4}\left(
\tilde{\eta}_{\alpha,k}^{2}\left( -c_{\alpha}+\tfrac{3}{2}\right)
+\left(
\tilde{\varsigma}_{\alpha,kk}-\frac{\tilde{\eta}_{\alpha,k}}{\upsilon_{k}%
}\right) \left( -c_{\alpha}+\tfrac{1}{2}\right) \right)
\nonumber\label{mkk}\\
& +\tfrac{1}{32\pi^{2}}\log\frac{m_{h_{1}}^{2}}{\Lambda^{2}}{\sum_{\alpha,k}%
}n_{\alpha}m_{\alpha}^{4}\left(
\tilde{\eta}_{\alpha,k}^{2}+\tilde{\varsigma
}_{\alpha,kk}-\frac{\tilde{\eta}_{\alpha,k}}{\upsilon_{k}}\right) .
\end{align}
By using (\ref{mkk}), the scale dependance in (\ref{Tri}) can be removed
straightforward. In the $\overline{DR}$ scheme ($c_{\alpha}=3/2$), the Higgs
triple couplings can be written as%
\begin{align}
\lambda_{ijl}^{(3)} & =\lambda_{ijl}^{(3-tree)}+\sum_{\alpha}\tfrac
{n_{\alpha}m_{\alpha}^{4}}{32\pi^{2}}\left[ \left( \xi_{\alpha,ijk}%
+\eta_{\alpha,i}\varsigma_{\alpha,jk}+\eta_{\alpha,j}\varsigma_{\alpha
,ik}+\eta_{\alpha,k}\varsigma_{\alpha,ij}\right) \log\tfrac{m_{\alpha}^{2}%
}{m_{h_{1}}^{2}}\right. \nonumber\\
& \left.
+\eta_{\alpha,i}\eta_{\alpha,j}\eta_{\alpha,k}+\eta_{\alpha
,i}\varsigma_{\alpha,jk}+\eta_{\alpha,j}\varsigma_{\alpha,ik}+\eta_{\alpha
,k}\varsigma_{\alpha,it}\right] \nonumber\\
& +\frac{A}{C}{\sum_{\alpha}}n_{\alpha}m_{\alpha}^{4}\left(
\xi_{\alpha
,ijk}+\eta_{\alpha,i}\varsigma_{\alpha,jk}+\eta_{\alpha,k}\varsigma
_{\alpha,ij}+\eta_{\alpha,j}\varsigma_{\alpha,ik}\right) ,\label{EP}\\
A & ={\sum_{k}}(m_{h_{k}}^{2}-\tfrac{\lambda_{k}}{3}\upsilon_{k}^{2}%
)-{\sum_{\alpha,k}}\tfrac{n_{\alpha}\tilde{\eta}_{\alpha,k}^{2}m_{\alpha}^{4}%
}{32\pi^{2}}-{\sum_{\alpha,k}}\tfrac{n_{\alpha}m_{\alpha}^{4}\left(
\tilde{\eta}_{\alpha,k}^{2}+\tilde{\varsigma}_{\alpha,kk}-\frac{\tilde{\eta
}_{\alpha,k}}{\upsilon_{k}}\right) }{32\pi^{2}}\log\tfrac{m_{\alpha}^{2}%
}{m_{h_{1}}^{2}}\nonumber\\
C & ={\sum_{\alpha,k}}n_{\alpha}m_{\alpha}^{4}\left( \tilde{\eta}%
_{\alpha,k}^{2}+\tilde{\varsigma}_{\alpha,kk}-\frac{\tilde{\eta}_{\alpha,k}%
}{\upsilon_{k}}\right) .
\end{align}
In our model, we have $u_{11}=u_{22}=c$ and $u_{12}=-u_{21}=s$. Then, the
coefficients in (\ref{ep}) for gauge bosons, top quark and $S_{0}$ scalar are%
\begin{align}
\tilde{\eta}_{W,1} & =\tilde{\eta}_{Z,1}=\tilde{\eta}_{t,1}=\frac
{2}{\upsilon},~\tilde{\eta}_{S_{0},1}=\frac{\lambda_{0}\upsilon}{m_{0}^{2}%
},~\tilde{\eta}_{S_{0},2}=\frac{\eta_{01}\upsilon_{1}}{m_{0}^{2}}.\nonumber\\
\tilde{\varsigma}_{W,11} &
=\tilde{\varsigma}_{Z,11}=\tilde{\varsigma
}_{t,11}=\frac{2}{\upsilon^{2}},~\tilde{\varsigma}_{S_{0},11}=\frac
{\lambda_{0}}{m_{0}^{2}},~\tilde{\varsigma}_{S_{0},22}=\frac{\eta_{01}}%
{m_{0}^{2}},
\end{align}
and all other parameters are vanishing. For the two CP-even scalars $h_{1,2}$,
we have
\begin{align}
\tilde{\eta}_{(1,2),1} & =\{\lambda+\lambda_{1}\mp\lbrack(\lambda
-\lambda_{1})\left( a-b\right)
+8\lambda_{1}^{2}\upsilon_{1}^{2}]/[2\left(
m_{2}^{2}-m_{1}^{2}\right) ]\}\upsilon/m_{1,2}^{2},\nonumber\\
\tilde{\eta}_{(1,2),2} & =\{\eta_{1}+\lambda_{1}\mp\lbrack(\lambda_{1}%
-\eta_{1})\left( a-b\right) +8\lambda_{1}^{2}\upsilon^{2}]/[2\left(
m_{2}^{2}-m_{1}^{2}\right) ]\}\upsilon_{1}/m_{1,2}^{2},
\end{align}
with%
\begin{equation}
a=-\mu^{2}+\lambda\upsilon^{2}/2+\lambda_{1}\upsilon_{1}^{2}/2,~b=-\mu_{1}%
^{2}+\lambda_{1}\upsilon^{2}/2+\eta_{1}\upsilon_{1}^{2}/2. \label{fm2}%
\end{equation}
While%
\begin{align}
\tilde{\varsigma}_{(1,2),11} & =\{\lambda+\lambda_{1}\mp\lbrack
(\lambda-\lambda_{1})\left( a-b\right) +(\lambda-\lambda_{1})^{2}%
\upsilon^{2}+8\lambda_{1}^{2}\upsilon_{1}^{2}]/[2\left( m_{2}^{2}-m_{1}%
^{2}\right) ]\pm\nonumber\\
& [\left( (\lambda-\lambda_{1})\left( a-b\right) \upsilon+8\lambda_{1}%
^{2}\upsilon\upsilon_{1}^{2}\right) ^{2}]/[4\left( m_{2}^{2}-m_{1}%
^{2}\right) ^{3}]\}/m_{1,2}^{2},\nonumber\\
\tilde{\varsigma}_{(1,2),12} & =\mp\lbrack\left( 15\lambda_{1}^{2}%
+\lambda\lambda_{1}-\lambda\eta_{1}+\lambda_{1}\eta_{1}\right)
\upsilon \upsilon_{1}]/[2m_{1,2}^{2}\left(
m_{2}^{2}-m_{1}^{2}\right) ]\pm
\lbrack((\lambda-\lambda_{1})\left( a-b\right) \upsilon\nonumber\\
& +8\lambda_{1}^{2}\upsilon\upsilon_{1}^{2})\left( (\lambda_{1}-\eta
_{1})\upsilon_{1}\left( a-b\right)
+8\lambda_{1}^{2}\upsilon^{2}\upsilon
_{1}\right) ]/[2m_{1,2}^{2}\left( m_{2}^{2}-m_{1}^{2}\right) ^{3}%
],\nonumber\\
\tilde{\varsigma}_{(1,2),22} &
=\{\eta_{1}+\lambda_{1}\mp\lbrack(\lambda
_{1}-\eta_{1})\left( a-b\right) +(\lambda_{1}-\eta_{1})^{2}\upsilon_{1}%
^{2}+8\lambda_{1}^{2}\upsilon^{2}]/[2\left(
m_{2}^{2}-m_{1}^{2}\right)
]\pm\nonumber\\
& [\left( (\lambda_{1}-\eta_{1})\left( a-b\right) \upsilon_{1}%
+8\lambda_{1}^{2}\upsilon^{2}\upsilon_{1}\right) ^{2}]/[4\left( m_{2}%
^{2}-m_{1}^{2}\right) ^{3}]\}/m_{1,2}^{2},
\end{align}
and%
\begin{align}
\tilde{\xi}_{(1,2),111} &
=\mp3\{(\lambda-\lambda_{1})^{2}\upsilon-[\left(
(\lambda-\lambda_{1})\left( a-b\right) \upsilon+8\lambda_{1}^{2}%
\upsilon\upsilon_{1}^{2}\right) ((\lambda-\lambda_{1})\left(
a-b\right)
+(\lambda-\lambda_{1})^{2}\upsilon^{2}\nonumber\\
& +8\lambda_{1}^{2}\upsilon_{1}^{2})]/[2\left(
m_{2}^{2}-m_{1}^{2}\right) ^{2}]+[\left( (\lambda-\lambda_{1})\left(
a-b\right) \upsilon+8\lambda
_{1}^{2}\upsilon\upsilon_{1}^{2}\right) ^{3}]/[4\left( m_{2}^{2}-m_{1}%
^{2}\right) ^{4}]\}\nonumber\\
& /\{2m_{1,2}^{2}\left( m_{2}^{2}-m_{1}^{2}\right) \},\nonumber\\
\tilde{\xi}_{(1,2),112} & =\mp\{\left(
15\lambda_{1}^{2}+\lambda\lambda
_{1}-\lambda\eta_{1}+\lambda_{1}\eta_{1}\right) \upsilon_{1}-[\left(
(\lambda_{1}-\eta_{1})\left( a-b\right) \upsilon_{1}+8\lambda_{1}%
^{2}\upsilon^{2}\upsilon_{1}\right) \times\nonumber\\
& \left( (\lambda-\lambda_{1})\left( a-b\right) +(\lambda-\lambda_{1}%
)^{2}\upsilon^{2}+8\lambda_{1}^{2}\upsilon_{1}^{2}\right)
+2((\lambda
-\lambda_{1})(\lambda_{1}-\eta_{1})\upsilon\upsilon_{1}\nonumber\\
& +16\lambda_{1}^{2}\upsilon\upsilon_{1})\left(
(\lambda-\lambda_{1})\left( a-b\right)
\upsilon+8\lambda_{1}^{2}\upsilon\upsilon_{1}^{2}\right)
]/[2\left( m_{2}^{2}-m_{1}^{2}\right) ^{2}]+\nonumber\\
& [3((\lambda_{1}-\eta_{1})\left( a-b\right) \upsilon_{1}+8\lambda_{1}%
^{2}\upsilon^{2}\upsilon_{1})\left( (\lambda-\lambda_{1})\left(
a-b\right)
\upsilon+8\lambda_{1}^{2}\upsilon\upsilon_{1}^{2}\right) ^{2}]\nonumber\\
& /[4\left( m_{2}^{2}-m_{1}^{2}\right) ^{4}]\}/\{2m_{1,2}^{2}\left(
m_{2}^{2}-m_{1}^{2}\right) \},\nonumber\\
\tilde{\xi}_{(1,2),122} & =\mp\{\left(
15\lambda_{1}^{2}+\lambda\lambda
_{1}-\lambda\eta_{1}+\lambda_{1}\eta_{1}\right) \upsilon-[\left(
(\lambda-\lambda_{1})\left( a-b\right) \upsilon+8\lambda_{1}^{2}%
\upsilon\upsilon_{1}^{2}\right) \times\nonumber\\
& \left( (\lambda_{1}-\eta_{1})\left( a-b\right) +(\lambda_{1}-\eta
_{1})^{2}\upsilon_{1}^{2}+8\lambda_{1}^{2}\upsilon^{2}\right)
+2((\lambda
-\lambda_{1})(\lambda_{1}-\eta_{1})\upsilon\upsilon_{1}\nonumber\\
& +16\lambda_{1}^{2}\upsilon\upsilon_{1})\left( (\lambda_{1}-\eta
_{1})\left( a-b\right)
\upsilon_{1}+8\lambda_{1}^{2}\upsilon^{2}\upsilon
_{1}\right) ]/[2\left( m_{2}^{2}-m_{1}^{2}\right) ^{2}]\nonumber\\
& +[3\left( (\lambda-\lambda_{1})\left( a-b\right) \upsilon+8\lambda
_{1}^{2}\upsilon\upsilon_{1}^{2}\right) \left(
(\lambda_{1}-\eta_{1})\left( a-b\right)
\upsilon_{1}+8\lambda_{1}^{2}\upsilon^{2}\upsilon_{1}\right)
^{2}]\nonumber\\
& /[4\left( m_{2}^{2}-m_{1}^{2}\right) ^{4}]\}/\{2m_{1,2}^{2}\left(
m_{2}^{2}-m_{1}^{2}\right) \},\nonumber\\
\tilde{\xi}_{(1,2),222} & =\mp3\{(\lambda_{1}-\eta_{1})^{2}\upsilon
_{1}-[\left( (\lambda_{1}-\eta_{1})\left( a-b\right) \upsilon_{1}%
+8\lambda_{1}^{2}\upsilon^{2}\upsilon_{1}\right) ((\lambda_{1}-\eta
_{1})\left( a-b\right) \nonumber\\
&
+(\lambda_{1}-\eta_{1})^{2}\upsilon_{1}^{2}+8\lambda_{1}^{2}\upsilon
^{2})]/[2\left( m_{2}^{2}-m_{1}^{2}\right) ^{2}]+[\left( (\lambda_{1}%
-\eta_{1})\left( a-b\right) \upsilon_{1}+8\lambda_{1}^{2}\upsilon
^{2}\upsilon_{1}\right) ^{2}]\nonumber\\
& /[4\left( m_{2}^{2}-m_{1}^{2}\right) ^{4}]\}/\{2m_{1,2}^{2}\left(
m_{2}^{2}-m_{1}^{2}\right) \}.
\end{align}

\end{document}